\documentclass[11pt,a4paper,fleqn]{article}
\usepackage{graphicx}
\usepackage[usenames,dvipsnames]{xcolor} 
\usepackage{amssymb,amsmath,mathtools,mathrsfs} 
\usepackage{epsfig,subfigure} 
\usepackage{booktabs,longtable} 
\usepackage{exscale,relsize} 
\usepackage[normalem]{ulem} 
\usepackage{enumerate} 
\usepackage{jcappub}
\usepackage{caption}
\usepackage{hyperref}
\usepackage{float}
\usepackage[utf8]{inputenc}
\usepackage{comment}
\usepackage{color,colortbl}

\makeindex

\begin{document}
\hypersetup{pageanchor=false} 
\title{A general theory of linear cosmological perturbations: bimetric theories}

\author[a,b]{Macarena Lagos,}
\author[a]{Pedro G. Ferreira}
\affiliation[a]{Astrophysics, University of Oxford, DWB,\\
Keble road, Oxford OX1 3RH, UK}
\affiliation[b]{Theoretical Physics, Blackett Laboratory, Imperial College London,\\
Prince Consort Road, London SW7 2BZ, UK}

\emailAdd{m.lagos13@imperial.ac.uk}
\emailAdd{p.ferreira1@physics.ox.ac.uk}

\abstract{We implement the method developed in \cite{Lagos:2016wyv} to construct the most general parametrised action for linear cosmological perturbations of bimetric theories of gravity. Specifically, we consider perturbations around a homogeneous and isotropic background, and identify the complete form of the action invariant under diffeomorphism transformations, as well as the number of free parameters characterising this cosmological class of theories. We discuss, in detail, the case without derivative interactions, and compare our results with those found in massive bigravity.}



\maketitle
\section{Introduction}
\label{sec:intro}
This paper is a follow-up of the work presented in \cite{Lagos:2016wyv}, in which we developed a method for constructing parametrised actions for linear cosmological perturbations for a given class of gravity theories, invariant under particular gauge symmetries. Our method allows us to describe a broad range of theories on cosmological scales in a unified manner, and enables us to ultimately test and compare the viability of gravity theories by constraining the parametrised actions with relevant observational data such as the Cosmic Microwave Background (CMB) or measures of Large Scale Structure (LSS) such as weak lensing and galaxy redshift surveys. Our proposal contributes to ongoing work by a number of groups who have focused on linear perturbation theory in scalar-tensor theories and other variants of modified gravity \cite{Creminelli:2008wc,Zuntz:2011aq,Baker:2012zs,Gubitosi:2012hu,Bloomfield:2012ff,Gleyzes:2013ooa,Bloomfield:2013efa,Gleyzes:2014rba,Battye:2012eu,Battye:2013ida,Skordis:2015yra}. 

The method in \cite{Lagos:2016wyv} is general and systematic and thus can be applied to a wide range of cases. In \cite{Lagos:2016wyv} we applied it to scalar-tensor and vector-tensor gravity theories. We showed that the form of the quadratic action, crucially, depends on the gauge transformation properties of any extra fields that may arise in a modified gravity theory, and for this reason the parametrised action can be very different depending on the field content of gravity. For instance, for diffeomorphism-invariant scalar-tensor theories there are only four free parameters determining the general form of the quadratic action for cosmological perturbations, whereas for vector-tensor theories there are ten. 

In order to extend previous work, and construct a parametrised model that spans as large a swathe of the landscape of gravitational theories as possible, in this paper we apply the method to diffeomorphism-invariant bimetric theories. We show that, around homogeneous and isotropic backgrounds, the most general quadratic action is determined by 29 free parameters, and propagates at most four scalar degrees of freedom (DoFs). However, if we do not allow derivative interactions between both metrics, the number of free parameters simplifies greatly, reducing to three. Furthermore, if we focus on actions that propagate only one DoF, the action has only two free parameters. Lately, bimetric theories have been studied extensively in a cosmological context \cite{Volkov:2011an,vonStrauss:2011mq,Comelli:2011zm,Capozziello:2012re,Comelli:2012db,Berg:2012kn,Akrami:2012vf,Sakakihara:2012iq,Akrami:2013ffa,Tamanini:2013xia,Aoki:2013joa,Akrami:2013pna,DeFelice:2013nba,Fasiello:2013woa,Volkov:2013roa,Konnig:2013gxa,Berezhiani:2013dw,DeFelice:2014nja,deRham:2014naa,Enander:2014xga,Konnig:2014dna,Solomon:2014dua,Comelli:2014bqa,Konnig:2014xva,Lagos:2014lca,Cusin:2014psa,Gumrukcuoglu:2014xba,Aoki:2014cla,Enander:2015vja,Nersisyan:2015oha,Soloviev:2015wya,Amendola:2015tua,Johnson:2015tfa,Konnig:2015lfa,Cusin:2015pya,Fasiello:2015csa,Akrami:2015qga,Aoki:2015xqa,Lagos:2015sya,Cusin:2015tmf,Comelli:2015pua,Heisenberg:2016spl, deRham:2010tw,D'Amico:2011jj,Gumrukcuoglu:2011ew,Gumrukcuoglu:2011zh,Vakili:2012tm,DeFelice:2012mx,Fasiello:2012rw,deRham:2014gla,Pereira:2015jua, Banados:2010ix,Scargill:2012kg,EscamillaRivera:2013hv,PhysRevD.89.024034,Bouhmadi-Lopez:2013lha,Cho:2014ija,Cho:2014jta,Cho:2014xaa,Jimenez:2015jqa}, and in this paper we show that we can recover the specific cases of massive bigravity \cite{deRham:2010ik, deRham:2010kj, Hassan:2011zd} and Eddington-inspired Born Infeld (EiBI) theory \cite{Vollick:2003qp, Banados:2010ix}, as bimetric theories encompassed by the parametrisation found. Due to the no-go theorem for ghost-free Lorentz-invariant bimetric theories with derivative interactions in \cite{deRham:2013tfa}, we devote most of the work in this paper to study non-derivative interactions, although we briefly discuss derivative interactions in the appendix.

This paper is structured as follows. In Section \ref{sec:method} we first review the method developed in \cite{Lagos:2016wyv}, and explain in detail how to implement it to bigravity theories, focusing on scalar perturbations. In Section \ref{sec:non-der} we present the results of the method in the specific case when the two metrics do not have any derivative interaction. We analyse the number of physical scalar DoFs propagating and the number of free parameters determining the general structure of this action. We also compare with massive bigravity and corroborate results found in previous works. Finally, in Section \ref{sec:discussion} we summarise and discuss the findings of this paper.

\section{Parametrising bimetric theories}
\label{sec:method}

In this section we summarise the method developed in \cite{Lagos:2016wyv} and then show how to apply it to bimetric theories. The objective of the method is to obtain a general, local, quadratic action for linear cosmological perturbations for a class of gravitational theories, with a given field content and gauge symmetries. This action will be expressed in terms of parameters --functions of the background-- in such a way that specific forms for the parameters lead to the action of a specific gravity theory. 
For simplicity, in what follows, we will be assuming a known form for the matter sector which couples to gravity, although it is straightforward to relax this assumption. We perform the following three steps to find the aforementioned parametrised action:

\begin{itemize}
\item[1.] {\bf Set up:} Assume a given number and type of fields present in the theory (gravity and matter). Define an ansatz for the cosmological background, and consider linear perturbations for each field around that background. Finally, choose a set of gauge symmetries that will leave the quadratic action invariant. 

\item[2.] {\bf General action:} Construct the most general local quadratic gravitational action, given the field content established in Step 1. Consider a set of perturbed building blocks $\delta \vec{\Theta}$, containing all the derivatives of the gravitational perturbation fields, up to a given maximum order. Use the building blocks to write down all possible quadratic terms, and construct the most general quadratic gravitational action $S_G^{(2)}$ by adding each one of these terms multiplied by an, a priori, free parameter (i.e.~unknown function of the background). The resulting gravitational action will contain all possible interactions between the perturbation fields. Finally, form the known matter content, calculate the quadratic matter action $S_m^{(2)}$ to get the total quadratic action $S_G^{(2)}+S_m^{(2)}$ determining the evolution of linear cosmological perturbations. 

\item[3.] {\bf Gauge invariance:} Find the most general linearly gauge-invariant quadratic action for perturbations. Consider the total action $S_G^{(2)}+ S_m^{(2)}$ from Step 2, and impose invariance under the desired gauge symmetries set in Step 1. In order to do this, find the Noether identities associated to the gauge symmetries: there will be one for each gauge parameter. Then, from each Noether identity, find the Noether constraints associated, which will, in general, be linear ordinary differential equations of the free parameters in $S_G^{(2)}$. Finally, solve the system of Noether constraints and replace the results in $S_G^{(2)}$. The resulting gravitational action will satisfy the Noether identities and, as consequence, be gauge-invariant under the desired symmetries. 
\end{itemize}

After performing these three steps, we will obtain the most general gravitational action for a class of theories, determined by a given field content and set of gauge symmetries, up to a maximum number of derivatives. From this result it is straightforward to identify the number of free parameters describing the linear cosmological evolution of the universe for these theories, and the number of physical, propagating, DoFs. In addition, from this action, we are able to constrain the free parameters with observational data and automatically translate these constraints into constraints on the gravitational theories. 

Before applying the steps to bimetric theories, we clarify that, for simplicity, we have slightly modified Step 2 compared to the one presented in \cite{Lagos:2016wyv}. As shown in \cite{Lagos:2016wyv}, in Step 2 the free parameters can be expressed as functional derivatives of the unknown underlying (non-linear) gravity theory. This is an interesting feature of the method as the parameters are related to characteristics of the fundamental theory, which is useful when building viable non-linear gravitational theories out of observational constraints on the parameters. However, in this paper we have skipped that part of Step 2, and thus we will not give such relations between the free parameters and the underlying gravity theory.


We now implement the previous method for bigravity theories. We show each one of the three steps in detail.\\

\noindent{\bf Step 1}: Consider a gravitational theory composed of two rank-2 tensor fields (or metrics) $g_{A\mu\nu}$ (with $A=\left\{1,2\right\}$), coupled to some additional matter fields. We focus on linear perturbations around a homogeneous and isotropic cosmological background. The tensor degrees of freedom arise from the following metrics:
\begin{equation}
g_{A\mu\nu}= \bar{g}_{A\mu\nu}+ \delta g_{A\mu\nu},
\end{equation}
where $\bar{g}_{A\mu\nu}$ describes the background of both metrics $A=1$ and $A=2$, assumed to be spatially-flat FRW metrics with line elements given by:
\begin{equation}
ds_1^2=-dt^2+a(t)^2d\vec{x}^2, \quad ds_2^2=-\bar{N}(t)^2dt^2+b(t)^2d\vec{x}^2.
\end{equation}
Here $a$ and $b$ are the scale factors of the metrics, and we have also introduced a non-trivial lapse term $\bar{N}$ for metric 2, as in general both metrics cannot be brought into the standard form with trivial shifts at the same time. Since we do not know what the underlying gravity theory is, the background functions $a$, $b$ and $\bar{N}$ are considered to be free functions of time. In addition, $\delta g_{A\mu\nu}$ are small first-order perturbations around the given background, which generally depend on space and time.

For simplicity and concreteness, we will assume that metric 1 is minimally coupled to a scalar field $\varphi$, which will represent our matter content. This means that metric 1 will be the physical metric describing the space-time, and therefore $a$ will be the scale factor of the expansion of the universe. The matter field can be expanded as follows:
\begin{equation}
\varphi =\bar{\varphi}(t)+\delta \varphi,
\end{equation}
where $\bar{\varphi}$ is the background value of the field, which has the same symmetries as the metrics, and thus depends on time only, whereas $\delta \varphi$ is the first-order perturbation of the field and, in general, depends on space and time. We remark that even though we use a minimally-coupled scalar field as matter content, we expect the same results for a general perfect fluid. Further generalisations could be made straightforwardly, such as coupling both metrics to matter through the composite metric proposed in \cite{deRham:2014naa,Noller:2014sta}, but such cases are left for future work.

We will be looking for actions which are quadratic in these perturbations and invariant under linear general coordinate transformations of the form $x^\mu\rightarrow x^\mu+\epsilon^\mu$, where $\epsilon^\mu$ is an arbitrary first-order perturbation of the coordinates $x^\mu$. Under these transformations the background stays the same, while the linear perturbations of the metrics $\delta g_{A\mu\nu}$ transform as (see \cite{Lagos:2016wyv} for an explicit derivation):
\begin{equation}
\delta g_{A\mu\nu}\rightarrow \delta g_{A\mu\nu} - {\bar g}_{A\mu\beta}\partial_\nu\epsilon^\beta - {\bar g}_{A\beta\nu}\partial_\mu\epsilon^\beta+\epsilon^\alpha {\bar g}_{A\mu\beta}{\bar g}_{A\nu\gamma} \left(\partial_\alpha {\bar g}^{A\beta\gamma}\right), \label{gauge:metric}
\end{equation}
whereas the scalar perturbation $\delta\varphi$ transforms as
\begin{eqnarray}\label{gauge:scalar}
\delta\varphi\rightarrow \delta\varphi-\dot{\bar{\varphi}}\pi.
\end{eqnarray}
Here, $\pi$ is an arbitrary parameter corresponding to the time component of the gauge parameter $\epsilon^\mu$, and the dot denotes a derivative with respect to the physical time $t$. Notice that, since we expect both metrics to be coupled, there will be only one copy of the diffeomorphism invariance, and thus both metrics transform with the same gauge parameter. \\

\noindent{\bf Step 2:} In this step we construct the most general local quadratic action for all the gravitational perturbation fields $\delta g_{A\mu\nu}$. This quadratic action will lead to linear equations of motion, assumed to have second-order derivatives at most.

Following the same notation as \cite{Lagos:2016wyv}, we use the Arnowitt-Deser-Misner (ADM) formalism, in which we separate space and time, and both metrics are decomposed into lapse functions $N_A$, shift functions $N^i_A$ and 3-dimensional spatial metrics $h_{Aij}$ in the following way:
\begin{equation}
g_{A00}=-{N^2_A}+h_{Aij}N^i_AN^j_A, \; g_{A0i}=h_{Aij}N_A^j, \; g_{Aij}=h_{Aij}. \label{ADMDecomposition} 
\end{equation}
Next, we write the set of perturbed building blocks $\delta \vec{\Theta}$, which includes all the perturbations the gravitational action can depend on. Specifically, this set will include all possible gravitational perturbations up to second-order derivatives for both metrics\footnote{We have ignored second-order time derivatives as they will be related -- via integration by parts -- to terms with first-order time derivatives.}:
\begin{align}\label{PertBlocks}
\delta \vec{\Theta}=&\Big( \delta N_A,\, \delta {\dot N}_A,\, \delta \partial_i N_A,\, \delta \partial_i\dot{N}_A,\, \delta \partial_i\partial_j N_A,\, \delta N^i_A,\, \delta {\dot N}^i_A, \delta \partial_j N^i_A, \, \delta \partial_j \dot{N}^i_A, \, \delta \partial_i\partial_j N^k_A, \nonumber\\
& \delta h_{Aij}, \, \delta \partial_i h_{Ajk},\, \delta \partial_i\partial_j h_{Akl}, \, \delta K^i_{A\phantom{i}j}\Big),
\end{align}
where the previous list includes all the terms for both subindices $A=\{1,2\}$. Here, we have replaced the time derivative terms $\delta \dot{h}_{Aij}$ of the spatial metrics by the extrinsic curvature tensors $\delta K^i_{A\phantom{i}j}$ -- this can always be done as there is a one-to-one relation between these two quantities. It is important to note that in \cite{Lagos:2016wyv} we also replaced second-order spatial derivatives of the spatial metrics by the 3-dimensional intrinsic curvature tensors $\delta R^i_{A\phantom{i}j}$, given that we were seeking linearly diffeomorphism-invariant actions, and thus spatial derivatives were expected to appear in that form. However, in this paper we do not use $\delta R^i_{A\phantom{i}j}$ because, in the case of bimetric theories, the interactions between the two metrics may cause the action to depend on a different combination of spatial derivatives, while still maintaining its diffeomorphism invariance. This is a new feature of the bimetric theories as in scalar-tensor and vector-tensor theories the spatial derivatives always appeared in the combination of $\delta R^i_{A\phantom{i}j}$.

Note that in eq.~(\ref{PertBlocks}) partial derivatives of the perturbation fields are taken with respect to the background metric of the corresponding perturbation field, and thus we raise and lower the indices of the perturbed building blocks with $\bar{h}_{Aij}$. Also, $\delta$ commutes with partial spatial derivatives and so, for instance, $\delta (\partial_i N_A) = \partial_i (\delta N_A)$. Finally, notice that even though the fields $g_{A\mu\nu}$ and $\varphi$ have only linear perturbations, the perturbed building blocks could have higher-order perturbations as result. Thus, we clarify that $\delta \vec{\Theta}$ contains both first and second-order perturbative terms; throughout this paper, however, $\delta$ refers only to first-order perturbations, unless stated otherwise. 

We now proceed to construct the most general quadratic bimetric action with second-order derivative equations of motion. In order to do that we write down all possible quadratic terms formed by the perturbed building blocks, and place an arbitrary parameter in front of each term. The resulting gravitational action can be written as follows:
\begin{eqnarray}
S_G^{(2)}=\int d^4 x \; \sum_{n=0}^2 {\cal L}_{T_1}^{n}+ {\cal L}_{T_2}^{n} + {\cal L}_{T_1T_2}^n. \label{GRaction}
\end{eqnarray}
Here, for a given $n$, ${\cal L}_{T_1}^{n}$ is the quadratic Lagrangian for the self-interactions of tensor 1, leading to $n$-order derivatives in the equations of motion. ${\cal L}_{T_2}^{n}$ is defined analogously for tensor 2, and finally ${\cal L}_{T_1T_2}^n$ is defined analogously for the interactions between both tensors.
We have the following Lagrangians for the self-interactions:
\begin{align}\label{LT0}
 {\cal L}_{T_A}^0&= \sqrt{-\bar{g}_A}\left[ \frac{1}{2}L_{ASS}\bar{h}_{Aij}\delta N^i_A\delta N^j_A +\frac{1}{2}L_{ANN}\left(\delta N_A\right)^2 + L_{AN} \left(\delta_2 N_A+\delta N_A \frac{ \delta\sqrt{h_A}}{\sqrt{-\bar{g}_A}}\right) \right. \nonumber\\
& \left. +L_{ANh}\delta N_A \delta h_A + \frac{1}{2}L_{Ahh+}\left(\delta h_A\right)^2 + L_{Ahh\times }\delta h^i_{Aj}\delta h^j_{Ai}\right] + \bar{N}_A\bar{L}_A\delta_2\sqrt{h_A},
\end{align}
\begin{align}\label{LT1}
{\cal L}_{T_A}^1& =\sqrt{-\bar{g}_A}\left[ L_{A\partial Sh+}\partial_i \delta N^i_A \delta h_A + 2L_{A\partial Sh\times } \delta h_{Aij}\partial^i \delta N^j_A + 2L_{AhK\times} \delta h^j_{Ai} \delta K^i_{Aj} \right.\nonumber\\
& \left. + L_{AhK+} \delta K_A\delta h_A+ L_{ANK}\delta N_A \delta K_A + L_{AN\partial S}\delta N_A \partial_i \delta N^i_A \right] ,
\end{align}
\begin{align}\label{LT2}
 {\cal L}_{T_A}^2&= \sqrt{-\bar{g}_A}\left[ L_{Ah\partial^2 h\times} \delta h_A\partial^i\partial^j\delta h_{Aij} + L_{Ah\partial^2h+ }\delta h_A \partial^2\delta h_A + L_{Ah\partial^2h\odot}\delta h_{Akl}\partial^k\partial^j\delta h_{Aij}\bar{h}^{li}_A \right. \nonumber\\
&+ 2L_{A\partial \dot{S}h\times }\delta h_{Aij}\partial^i \delta \dot{N}^j_A + L_{A\partial \dot{S}h+}\partial_i \delta \dot{N}^i_A\delta h_A + \frac{1}{2}L_{A\partial S\partial S+} \left(\partial_i \delta N^i_A\right)\left(\partial_j \delta N^j_A\right) \nonumber\\
& + \frac{1}{2}L_{A\dot{S}\dot{S}}\bar{h}_{ij} \delta \dot{N}^j_A \delta \dot{N}^i_A + \frac{1}{2}L_{A\partial S\partial S \times}\bar{h}_{lj}\left( \partial_i \delta N^l_A\right)\left(\partial^i\delta N^j_A\right) + \frac{1}{2}L_{AKK+}\left(\delta K_A\right)^2  \nonumber\\
&+ L_{AKK\times}\delta K^{i}_{Aj} \delta K^j_{Ai} + \frac{1}{2}L_{A\dot{N}\dot{N}}\left(\delta \dot{N}_A\right)^2 + \frac{1}{2}L_{A\partial N \partial N} \partial^i \delta N_A \partial_i\delta N_A \nonumber \\
& + L_{Ah\partial^2N+}\delta h_A\partial^2 \delta N_A + 2L_{Ah\partial^2N\times}\delta h_{Aij}\partial^i\partial^j \delta N_A + L_{AK\dot{N}}\delta K_A\delta \dot{N}_A \nonumber\\
& \left. +L_{AN\partial \dot{S}} \left(\partial_j\delta \dot{N}^j_A\right) \delta N_A + L_{A\partial S K+}\delta K_A\partial_i \delta N^i_A + 2L_{A\partial S K\times } \delta K^i_{Aj}\partial_i \delta N^j_A \right].
\end{align}
Here, the $L$ parameters are free functions of time with a subscript indicating the type of self-interaction they determine. Also, $\sqrt{-\bar{g}_A}$ correspond to the square root of the determinant of the 4-dimensional background metrics $\bar{g}_{A\mu\nu}$. We remind the reader that spatial indices are raised and lowered with the 3-dimensional background metrics, so in the previous actions we have simplified notation by introducing the perturbation fields $\delta h_A$, which are given by $\delta h_A=\bar{h}^{ij}\delta h_{Aij}$. Also, notice that we are using $\delta_2$ to describe quadratic perturbations \footnote{We have ignored the term $\delta_2 K^i_{\phantom{i}j}$ as it is related to other terms in the previous action (see \cite{Lagos:2016wyv}).}. 

In eq.~(\ref{LT0}) we have introduced the perturbations $\delta \sqrt{h_A}$ and $\delta_2 \sqrt{h_A}$, which are the linear and quadratic perturbations of the square root of the determinant of the 3-dimensional metrics. These perturbations are related to those of the building blocks by $\delta \sqrt{h_A}= \frac{1}{2}\sqrt{\bar{h}_A}\delta h_A$ and $\delta_2 \sqrt{h_A}= \frac{1}{8}\sqrt{\bar{h}_A}(\delta h_A)^2-\frac{1}{4}\sqrt{\bar{h}_A}\delta h^{i}_{Aj}\delta h^{j}_{Ai}$, where $\sqrt{\bar{h}_A}$ are the square root of the determinant of the 3-dimensional background metrics $\bar{h}_{Aij}$. As we will explain later on, we have introduced these perturbations because they will cancel out terms of the exact same form in the quadratic matter action. 

We also show explicitly the interaction terms between both metrics:
\begin{align}\label{LT1T20}
 {\cal L}_{T_1T_2}^{0}&= P_{N_2 h_1}\delta N_2 \delta h_1 + P_{N_2 N_1 } \delta N_1 \delta N_2 + P_{ S_2 S_1} \delta N^j_2 \delta N^i_1\bar{h}_{1ij}+ P_{ h_1 h_2+} \delta h_1 \delta h_2 \nonumber\\
& +P_{ h_1 h_2\times} \bar{h}^{ik}_1 \bar{h}^{jl}_1\delta h_{1ij}\delta h_{2kl} + P_{h_2 N_1}\delta h_2 \delta N_1,
\end{align}
\begin{align}\label{LT1T21}
{\cal L}_{T_1T_2}^{1}&= P_{\dot{N}_2 h_1 }\delta \dot{N }_2 \delta h_1 +P_{N_2\partial S_1} \delta N_2 \partial_i\delta N^i_1 + P_{\dot{N}_2N_1 } \delta \dot{N }_2\delta N_1 + P_{ S_2 \partial h_1+ }\partial_i \delta h_1 \delta N^i_2 \nonumber\\
& + 2P_{ S_2 \partial h_1\times}\partial^i \delta h_{1ij}\delta N^j_2 + P_{\dot{S}_2S_1}\delta N^i_1 \dot{N}^j_2 \bar{h}_{1ij} + P_{N_1\partial S_1 }\delta N_1 \partial_i \delta N^i_2 + P_{h_2 K_1+} \delta K_1 \delta h_2 \nonumber\\
& + P_{h_2 K_1\times } \bar{h}^{ik}_1\delta K^j_{i1} \delta h_{2jk} + P_{K_2 N_1}\delta K_2 \delta N_1 + P_{h_2 \partial S_1 + } \delta h_2 \partial_i\delta N^i_1 \nonumber\\
&+ P_{h_2 \partial S_1 \times } \delta h_{2ij} \partial^i\delta N^j_1,
\end{align}
\begin{align}\label{T1T22}
{\cal L}_{T_1T_2}^{2}&= P_{h_1\partial^2 N_2+}\delta h_1 \partial^2\delta N_2+ P_{h_1\partial^2 N_2\times}\delta h_{1ij} \partial^i\partial^j\delta N_2 + P_{K_1\dot{N}_2}\delta K_1 \delta \dot{N}_2  \nonumber\\
& + P_{\dot{N}_1\dot{N}_2}\delta\dot{N}_1 \delta \dot{N}_2 + P_{\dot{N}_2\partial S_1}\partial_i\delta N^i_1\delta \dot{N}_2 + P_{ N_2\partial^2 N_1 }\delta N_2\partial^2\delta N_1 + P_{\partial S_2 \dot{N}_1 } \delta \dot{N}_1 \partial_i \delta N^i_2 \nonumber\\
& + P_{ \dot{S}_1 \dot{S}_2} \bar{h}_{1ij}\delta \dot{N}^i_1 \delta \dot{N}^j_2 + P_{\partial S_2 K_1+} \delta K_1\partial_i \delta N^i_2 + 2P_{\partial S_2 K_1\times}\delta K^i_{j1}\partial_i \delta N^j_2 \nonumber\\
 & + P_{\partial S_1 \partial S_2 +}\partial_i \delta N^i_1 \partial_j \delta N^j_2 + P_{\partial S_1 \partial S_2 \times }\bar{h}_{1kj}\partial^i \delta N^k_1 \partial_i \delta N^j_2 + P_{K_2 \dot{N}_1}\delta K_2 \delta \dot{N}_1\nonumber\\
& + P_{K_1K_2+} \delta K_1 \delta K_2 + P_{K_1K_2\times} \delta K^i_{j1} \delta K^j_{i2}+ P_{h_2 \partial^2 N_1+}\delta h_2 \partial^2 \delta N_1 \nonumber\\
& + P_{h_2 \partial^2 N_1\times }\delta h_{2ij} \partial^i\partial^j \delta N_1 + P_{K_2 \partial S_1+}\delta K_2 \partial_i \delta N^i_1 + P_{K_2 \partial S_1\times}\delta K^i_{j2} \partial_i \delta N^j_1 \nonumber\\
&+ P_{h_2 \partial^2 h_1 +}\delta h_2 \partial^2 \delta h_1 + P_{h_2 \partial^2 h_1 \times 1}\delta h_2 \partial^i\partial^j \delta h_{1ij}+ P_{h_2 \partial^2 h_1 \times 2}\delta h_{2ij} \partial^i\partial^j \delta h_{1}\nonumber\\
& + P_{h_2 \partial^2 h_1 \odot}\delta h_{2il} \partial^i\partial^j \delta h_{1jk}\bar{h}^{lk},
\end{align}
where the $P$ parameters are free functions of time with a subscript indicating the type of interaction they determine. We clarify that $\partial^2=\partial^i\partial_i$ where, in general, the derivatives acting on a given field have a index that is raised or lowered with the background metric of that given field. For instance, the term $\delta N_2 \partial^2 \delta N_1$ can be equivalently expressed as:
\begin{equation}
\delta N_2 \partial^2 \delta N_1 = \delta N_2 \left(\partial_i\partial_j\bar{h}^{ij}_1\delta N_1\right)=\frac{\delta^{ij}}{a^2} \delta N_2 \partial_i\partial_j\delta N_1.
\end{equation}
Finally, we note that in all these Lagrangians (self-interactions and interactions between both metrics) we have integrated by parts, and written only the independent terms, so we express the action in terms of a minimal set of parameters. 

Now that we have written the most general quadratic action for two metrics, leading to second-order derivative equations of motion, we calculate the total quadratic action by adding the matter contribution. As previously mentioned, we will assume that metric 1 is minimally coupled to a scalar field $\varphi$, and thus the matter action has the following form:
\begin{equation}
S_m=-\int d^4x \; \sqrt{-g_1}\left( \frac{1}{2}\partial_\mu \varphi \partial^\mu \varphi + V(\varphi)\right),
\end{equation}
where $V(\varphi)$ is some potential, and the indices of the partial derivatives are raised and lowered using $g_{1\mu\nu}$. This action can be straightforwardly written in terms of the ADM variables, and Taylor expanded up to second order in the linear perturbations of the metric and scalar field to get:
\begin{align}\label{SecondSmv2}
S_m^{(2)}&= -\int d^4x\left\{-P_m\delta_2 \sqrt{h_1}+a^3\rho_m\left(\delta_2N_1+\delta N_1 \frac{\delta\sqrt{h_1}}{a^3}\right)-\frac{a^3}{2}\left(P_m+\rho_m\right)\left(\delta N_1\right)^2 \right. \nonumber \\ 
& \left.+ a^3\left[ \frac{1}{2}\bar{V}^{''}\delta\varphi^2 + \left(\bar{V}'\delta\varphi + \dot{\bar{\varphi}}\delta\dot{\varphi}\right) \delta N_1+ \dot{\bar{\varphi}}\partial_i\delta \varphi\delta N_1^i -\frac{1}{2}\delta\dot{\varphi}^2 + \frac{1}{2}\bar{h}_1^{ij}\partial_j\delta\varphi\partial_i\delta \varphi \right]\right. \nonumber \\ 
& \left.+\delta\sqrt{h_1}\left( \bar{V}'\delta\varphi - \delta\dot{\varphi}\dot{\bar{\varphi}}\right)\right\},
\end{align}
where $\bar{V}$ is the potential evaluated at the background value $\bar{\varphi}$, and the primes denote derivatives with regards to the scalar field. Also, we have introduced the background energy density $\rho_m$ and pressure $P_m$ of the fluid, given by: 
\begin{equation}
\rho_m=\frac{1}{2}\dot{\bar{\varphi}}^2+\bar{V}, \; P_m=\frac{1}{2}\dot{\bar{\varphi}}^2-\bar{V}.
\end{equation}
We note that $S_m^{(2)}$ does not only contain linear and quadratic terms on the perturbations of the matter field, but also quadratic terms in the perturbations of metric 1. In fact, the first two terms of the quadratic matter action in eq.~(\ref{SecondSmv2}) will cancel out exactly the similar two terms for metric 1 in eq.~(\ref{LT0}). This is because the coefficients $L_{AN}$ and $\bar{L}_A$ are not arbitrary, and they are defined in such a way that: 
\begin{equation}\label{2ParamsRelations}
L_{1N}=\rho_m, \; \bar{L}_1=-P_m, \; L_{2N}=0,\; \bar{L}_2=0.
\end{equation}
These relations are background equations and come from the linear expansion of the underlying gravity theory. For simplicity, we do not perform that expansion in this paper but we refer the reader to \cite{Lagos:2016wyv} for the derivation of these relations.

Later on, it will be of importance the fact that the matter action also leads to the following background equation:
\begin{equation}
\ddot{\bar{\varphi}}+3H\dot{\bar{\varphi}}+\bar{V}'=0, \label{N4}
\end{equation} 
where $H=\dot{a}/a$ is the Hubble rate. This equation acts as a constraint on the, a priori, free background functions $a$ and $\bar{\varphi}$.

Finally, we can see that the total quadratic action $S^{(2)}=S^{(2)}_m+S^{(2)}_G$ contains 99 free parameters $L$s and $P$s\footnote{Here we are not counting the coefficients $L_{AN}$ and $\bar{L}$, as their values are given in eq.~(\ref{2ParamsRelations}).}, and it is the most general quadratic action that can be written for linear perturbations of two metrics around a spatially-flat FRW background, with a minimal coupling to a scalar field, and leading to second-order derivative equations of motion. In addition, we have four background functions $a$, $b$, $\bar{N}$ and $\bar{\varphi}$ determining the background evolution, and thus affecting the linear perturbations, from which only three are independent due to the constraint of eq.~(\ref{N4}). We remark that, contrary to most known bigravity theories, we have assumed that the perturbations $\delta N_{A}$ and $\delta N_A^i$ are dynamical fields (have time-derivatives) and we will let the Noether identities dictate the consistent kinetic structure of these terms in a gauge-invariant action.\\

\noindent{\bf Step 3:} We now proceed to impose diffeomorphism invariance on the total quadratic action $S^{(2)}$ obtained in Step 2. First, we follow the standard Scalar-Vector-Tensor (SVT) decomposition \cite{1992PhR...215..203M} of perturbations. Since scalar, vector and tensor perturbations evolve independently at the linear level in a homogeneous and isotropic background, we can study each one separately. In this paper, we only analyse scalar perturbations, as they are the seeds of large-scale structure in the energy density field, and therefore cosmologically relevant. In this case, the line element of both metrics (including the background and linear perturbations) will be expressed as: 
\begin{align}
&ds_1^2=-\left(1+\Phi_1\right)dt^2+2\partial_i B_1 dt dx^i +a^2\left[\left(1-2\Psi_1\right)\delta_{ij}+2\partial_i\partial_j E_1\right]dx^i dx^j,\label{ScalarPertMetric1}\\
&ds_2^2=-\bar{N}^2\left(1+\Phi_2\right)dt^2+2\bar{N}\partial_i B_2 dt dx^i +b^2\left[\left(1-2\Psi_2\right)\delta_{ij}+2\partial_i\partial_j E_2\right]dx^i dx^j,\label{ScalarPertMetric2}
\end{align}
where the fields $\Phi_A$, $B_A$, $\Psi_A$ and $E_A$ for $A=\{1,2\}$ are the first-order scalar perturbations for both metrics $g_{A\mu\nu}$. From eq.~(\ref{ScalarPertMetric1})-(\ref{ScalarPertMetric2}) we can find the perturbed ADM variables at linear order:
\begin{align}
&\delta N_1= \Phi_1, \; \delta N^i_1 = \bar{h}^{ij}_1 \partial_j B_1, \; \delta h_{1ij}= a^2\left[-2\Psi_1\delta_{ij}+2\partial_i\partial_j E_1\right],\label{ADMmetric1}\\
&\delta N_2= \bar{N}\Phi_2, \; \delta N^i_2 = \bar{N}\bar{h}^{ij}_2 \partial_j B_2, \; \delta h_{2ij}= b^2\left[-2\Psi_2\delta_{ij}+2\partial_i\partial_j E_2\right],\label{ADMmetric2}
\end{align} 
as well as the rest of the perturbed building blocks. See Appendix \ref{app:metricperts} for a list of relevant quantities, that appear in the gravitational quadratic action $S_G^{(2)}$, in terms of the scalar perturbations. In these equations, $\bar{h}^{ij}_A$ are the inverse tensors of the 3-dimensional spatial background metrics.

We can now express the total quadratic action in terms of the 8 scalar perturbations, and the matter perturbation field $\delta \varphi$. Note that even though we only allow up to two derivatives of the metric perturbations $\delta N$, $\delta N^i$ and $\delta h_{ij}$, this means that we will have higher-order derivatives of the scalar perturbations. Next, we impose that the action is invariant under the linear diffeomorphism transformations given in eq.~(\ref{gauge:metric})-(\ref{gauge:scalar}). In terms of the scalar perturbations, the transformation of the metric perturbations is the following: 
 \begin{eqnarray}
 \tilde{\Phi}_1&=&\Phi_1 -{\dot \pi}, \quad \tilde{B}_1=B_1+{\pi}-a^2{\dot \epsilon},\quad \tilde{\Psi}_1=\Psi_1+ H{\pi}, \quad \tilde{E}_1=E_1 -\epsilon,\label{gauge:scalarsmetric1}\\
 \tilde{\Phi}_2&=&\Phi_2 -{\dot \pi}-H_N\pi, \quad \tilde{B}_2=B_2+\bar{N}{\pi}-\frac{b^2}{\bar{N}}{\dot \epsilon},\quad \tilde{\Psi}_2=\Psi_2+ H_b{\pi}, \quad \tilde{E}_2=E_2 -\epsilon, \label{gauge:scalarsmetric2}
 \end{eqnarray}
where we have defined $H=\dot{a}/a$, $H_b=\dot{b}/b$ and $H_N=\dot{\bar{N}}/\bar{N}$. Note that we have also used the SVT decomposition for the gauge parameter $\epsilon^\mu$, and written its two scalar components $\pi$ and $\epsilon$ such that $\epsilon^\mu=(\pi,\delta^{ij}\partial_j\epsilon)$. 

In order that the action is diffeomorphism invariant, we need to determine the Noether identities that arise for {\it both} spatial and temporal linear gauge transformations. These identities are the constraints which, when enforced, make the action gauge invariant. We calculate the Noether identities by first taking an infinitesimal variation of the total quadratic action with regards to each one of the scalar perturbations. This variation can be written as:
\begin{equation}
\hat{\delta} S^{(2)}= \hat{\delta} S_G^{(2)}+\hat{\delta} S_m^{(2)}=\int d^4 x\; \left[{\cal E}_{\Phi_A}\hat{\delta} \Phi_A+{\cal E}_{B_A}\hat{\delta} B_A+{\cal E}_{\Psi_A}\hat{\delta}\Psi_A+{\cal E}_{E_A}\hat{\delta} E_A+ {\cal E}_{\delta \varphi}\hat{\delta} \left(\delta \varphi\right) \right] \label{epsilonscalars}
\end{equation}
where ${\cal E}_X$ is the equation of motion of the perturbation field $X$, and $\hat{\delta}$ stands for functional variation. Here, there is an implicit sum over the subindex $A$. We then replace the variations of the fields by the corresponding gauge transformations in eq.~(\ref{gauge:scalar}) and eq.~(\ref{gauge:scalarsmetric1})-(\ref{gauge:scalarsmetric2}), and integrate by parts to end up with: 
\begin{eqnarray}
\hat{\delta}{g} S^{(2)} &=&\int d^4x\; \left[{\cal E}_{B_1}+\bar{N}{\cal E}_{B_2}+H{\cal E}_{\Psi_1}+H_b{\cal E}_{\Psi_2}+{\dot {\cal E}}_{\Phi_1}+ {\dot {\cal E}}_{\Phi_2}-H_N {\cal E}_{\Phi_2}- {\cal E}_{\delta\varphi} \dot{\bar{\varphi}}\right]\pi \nonumber \\
 &+&\int d^4x\; \left[-{\cal E}_{E_1}-{\cal E}_{E_2}+\frac{d}{dt}\left(a^2{\cal E}_{B_1}\right)+\frac{d}{dt}\left(\frac{b^2}{\bar{N}}{\cal E}_{B_2}\right)\right]\epsilon,
\end{eqnarray}
where the expression $\hat{\delta}_g$ stands for the functional variation of the action due to the gauge transformation. Given that the total quadratic action should be invariant under these gauge transformations, and given that both $\pi$ and $\epsilon$ are arbitrary and independent, each set of brackets should be zero; this gives us two \textit{Noether identities}, one associated to each scalar gauge parameter. Furthermore, each combination of free coefficients, inside each of the brackets, multiplying the perturbation fields and their derivatives such as $\Phi_A$, $ {\dot \Phi}_A$, $\partial^2 \Phi_A$, $\Psi_A$, etc, must be individually zero for the Noether identities to be satisfied off-shell. This gives a set of \textit{Noether constraints}.
As previously mentioned, these constraints will be, in general, linear ordinary differential equations for the coefficients $L$s and $P$s. However, for the bimetric case presented in this paper, these Noether constraints can be solved algebraically. We solve all of these constraints and replace the solutions in the quadratic action, resulting in an action satisfying the Noether identities and, as consequence, gauge invariant. Therefore, the resulting action will be the most general linearly diffeomorphism-invariant local quadratic action, given the field content. 

It is important to remark that even though we only impose gauge invariance under the scalar gauge parameters $\pi$ and $\epsilon$, the resulting action will also be gauge-invariant under the vector gauge parameter $\epsilon^{Ti}$, and thus it will be fully invariant under linear coordinate transformations. In turn, this means that the independent free parameters characterising the action for scalar perturbations are a superset of those characterising the vector and tensor perturbations.


The resulting gauge-invariant bimetric action is lengthy so we do not show it explicitly here but we mention some general characteristics. We find that the final action depends only on 29 free parameters, in addition to the four background functions $a$, $b$, $\bar{N}$ and $\bar{\varphi}$, which add three independent free functions due to the constraint in eq.~(\ref{N4}). Therefore, there are $29+3$ free parameters determining the background and linear cosmological evolution of these bimetric theories. In Appendix \ref{App:parameters} we give expressions for all these parameters in terms of the coefficients $L$s and $P$s, and we connect them with the well-known parameters present in the scalar-tensor parametrisation EFT of dark energy \cite{Gleyzes:2014rba}. 

In addition, we find that in the final action the eight metric scalar fields $\Phi_A$, $B_A$, $\Psi_A$ and $E_A$, have dynamical terms, i.e.~time derivatives. However, the fields $\Phi_A$ and $B_A$ appear in specific combinations such that if we introduce two new fields $\Phi_3$ and $B_3$:
\begin{equation}\label{Phi3B3}
\Phi_1=\Phi_3+\Phi_2, \quad B_1=B_3+ \frac{a^2\bar{N}}{b^2}B_2, 
\end{equation}
then $\Phi_3$ and $B_3$ appear as dynamical fields whereas $\Phi_2$ and $B_2$ become auxiliary fields (without time derivatives). This means that the final gravitational action propagates at most four physical scalar DoFs. The counting goes as follows: there are two auxiliary variables that can be worked out from their own equations of motion, and therefore expressed entirely in terms of the 6 remaining dynamical fields. In addition, we have two scalar gauge parameters that we can use to fix the gauge and eliminate two dynamical fields. Therefore, the final action has at most four physical scalar DoFs, although for specific values of the parameters (and background evolutions) there could be less. 

From our results it is not possible to know where the four scalar DoFs are coming from (e.g.~massive gravitons or ghosts), but we do know that all well-known healthy bimetric theories propagate at most one scalar DoF, signalling the possible presence of unstable modes in the action found in this paper. This shows a crucial feature of any approach based on linearized theories solely. The consistency of a full theory requires background, linearized perturbative and higher-order perturbative contributions all to be consistent, i.e.~to avoid the propagation of unstable degrees of freedom such as ghosts. And so, crucially, while all well-behaved theories will map onto the free functions in our linearized perturbation theory parametrisation, not all possible functional forms for these seemingly free functions are associated with healthy theories. This happens for the very simple reason that there is more to a full theory than the action it gives rise to for linear perturbations, and that there are additional constraints not captured by any formalism based on linearized perturbations. These extra constraints would reduce the free functions and the number of propagating fields we have found further. A detailed analysis on the construction of possible fundamental consistent theories leading to the parametrised bimetric action found here is beyond the scope of this paper, but it is certainly relevant and requires further work.

In the next section, we focus on the specific case when there are no derivative interactions between both metrics. This case is interesting as it encompasses most well-known bimetric theories such as massive bigravity and EiBI. Furthermore, there is a no-go theorem for the existence of ghost-free Lorentz-invariant derivative interactions \cite{deRham:2013tfa} for massive gravity, rendering the general case of derivative interactions likely to propagate unstable modes. Notwithstanding, we do briefly discuss derivative interactions in Appendix \ref{sec:der}, as they might still be relevant in the context of Lorentz-breaking theories.

\section{A reduced case: excluding derivative interactions}
\label{sec:non-der}
In this section we study the general structure of the parametrised bimetric action in the absence of derivative interactions between both metrics. The starting point is the general action we found in the previous section, which had 29 free parameters. In this action we impose that all derivative interactions vanish, i.e.~${\cal L}_{T_1T_2}^1={\cal L}_{T_1T_2}^2=0$, which enforces the relations on the free parameters that we present in Appendix \ref{App:ParamNonDer}. Specifically, we find 26 relations, reducing greatly the number of free parameters to only three. These three parameters are:
\begin{align}
M_{1}^2&=2L_{1KK\times},\\
\alpha_{L}&=-\frac{1}{2M_1^2H^2}\left(\bar{L}_1-4L_{1hh+}-8L_{1hh\times}\right), 
\end{align}
\begin{equation}
\alpha_{E}=-\frac{2}{M_2^2}\left[\bar{N}^2T_{2Nh}-H_b\left(2L_{2hK\times}+3L_{2hK+}\right)\right]. 
\end{equation}
Therefore, if we take into account the background functions, there are, in total, $3+3$ free functions of time determining the evolution of the background and linear perturbations in this subclass of theories: bimetric theories without derivative interactions. In Appendix \ref{App:ParamNonDer} we show the quadratic action without derivative interactions.

In this case we find that the four fields $B_{A}$ and $\Phi_{A}$ appear as auxiliary variables, i.e.~do not have any time derivatives. This means that the resulting gravitational action propagates at most two scalar DoFs. This result is consistent with previous analyses of massive gravity, where it has been shown that most potential interactions between two metrics lead to the propagation of a helicity-0 mode for a massive graviton and an extra unstable scalar mode, the Boulware-Deser ghost \cite{Boulware:1973my, ArkaniHamed:2002sp, Creminelli:2005qk}. 

In order to get a healthy action we construct actions that propagate only one scalar DoF (although there are some trivial healthy cases that propagate no scalar, as we will see later on). We do this by imposing that one of the dynamical fields is an auxiliary variable, i.e.~by setting to zero the coefficients of the kinetic terms of one field, after integrating out the four auxiliary fields $B_{A}$ and $\Phi_{A}$. As we will see later on, the resulting action is a generalisation of massive bigravity and thus the only propagating physical field should correspond to the helicity-0 mode of a massive graviton. We find that the only non-trivial situation we can have is when $\Psi_2$ becomes an auxiliary variable \footnote{In any other possible case the resulting action will lead to a copy of the linear Einstein-Hilbert action, and thus the gravitational action does not propagate any scalar DoF.}, which imposes one extra constraint on the parameters:
\begin{equation}\label{ConstraintOneDoF}
\alpha_E=\frac{H_b\left(\rho_m+P_m +2\dot{H}M_1^2\right)}{r^3(H-H_b)M_2^2},
\end{equation}
where we have introduced the scale factor ratio $r=b/a$, and the mass scale $M_2^2=2L_{2KK\times}$.
Therefore, the most general bimetric quadratic action without derivative interactions {\it and} propagating only one scalar DoF, depends on $2+3$ free functions of time. From now on, we focus on such a subclass of actions. The resulting parametrised action is much simpler in this case, and can be written in the following form: 
\begin{equation}\label{TotalActionNonDer}
S^{(2)}=S_{T_1}^{(2)}+ S_{T_2}^{(2)}+ S_{T_1T_2}^{(2)}+ S_{\varphi}^{(2)},
\end{equation}
where $S_{T_A}^{(2)}$ is the action for the self-interaction terms of the metric $g_{A\mu\nu}$, whereas $S_{T_1T_2}^{(2)}$ is the action for the interaction terms between both metrics, and $S_{\varphi}^{(2)}$ includes all the terms involving the matter perturbation $\delta \varphi$ in eq.~(\ref{SecondSmv2}). Notice that $S_{T_1}^{(2)}$ does include the quadratic terms of the metric perturbations coming from the matter action $S_m^{(2)}$. These actions are given by:

\begin{align}\label{S1NonDer}
S_{T_1}^{(2)}&=\int d^4x\; a^3M_1^2\left[-3\dot{\Psi}_1^2-6H\dot{\Psi}_1\Phi_1+2a^2\partial^2\dot{E}_1(\dot{\Psi}_1+H\Phi_1)-2\dot{\Psi}_1 \partial^2B_1 \right. \nonumber\\
& -\left(1+\frac{d\ln M_1^2}{d\ln a}\right)\Psi_1\partial^2 \Psi_1 +2\Phi_1\partial^2\Psi_1-2H\Phi_1\partial^2 B_1- \left(3H^2 - \frac{\dot{\bar{\varphi}}^2}{2M_1^2} \right)\Phi_1^2\nonumber\\
& + \frac{r^2Z}{2(\bar{N}+r)}\left(\partial^i B_1\right)\left(\partial_iB_1\right) + r \left(Z\frac{d\ln M_1^2}{d\ln a}+2\tilde{Z}\right) \Psi_1\left(\frac{3}{2}\Psi_1-a^2\partial^2E_1\right) \nonumber\\
& \left. +\frac{(\bar{N}-r)}{H-H_b}HZ\Phi_1\left(3\Psi_1- a^2\partial^2E_1\right) +a^4\alpha_{L}H^2\left(\partial^2E_1\right)^2 \right], 
\end{align}
\begin{align}\label{S2NonDer}
S_{T_2}^{(2)}&=\int d^4x\; \bar{N}b^3M_2^2\left[-3\frac{\dot{\Psi}_2^2}{\bar{N}^2}-6\frac{H_b}{\bar{N}^2}\dot{\Psi}_2\Phi_2+2\frac{b^2}{\bar{N}^2}\partial^2\dot{E}_2(\dot{\Psi}_2+H_b\Phi_2)-2\frac{\dot{\Psi}_2}{\bar{N}} \partial^2B_2 \right. \nonumber\\
& -\left(1+\frac{d\ln M_2^2}{d\ln b}\right)\Psi_2\partial^2 \Psi_2 +2\Phi_2\partial^2\Psi_2-2\frac{H_b}{\bar{N}}\Phi_2\partial^2 B_2-3\frac{H_b^2}{\bar{N}^2}\Phi_2^2 \nonumber\\
& + \frac{\nu^2\bar{N}Z}{2r^3(\bar{N}+r)}\left(\partial^i B_2\right)\left(\partial_iB_2\right)+ \frac{\nu^2}{\bar{N}r^2}\left(Z\frac{d\ln M_1^2}{d\ln a}+2\tilde{Z}\right) \Psi_2\left(\frac{3}{2}\Psi_2-b^2\partial^2E_2\right) \nonumber\\
& \left. + H_b\nu^2Z\frac{(\bar{N}-r)}{\bar{N}r^3(H-H_b)}\Phi_2\left(3\Psi_2- b^2\partial^2E_2\right) +\frac{r}{\bar{N}}\alpha_{L}a^4H^2\nu^2\left(\partial^2E_2\right)^2 \right],
\end{align}
\begin{align}\label{ST1T2NonDer}
S_{T_1T_2}^{(2)}&=\int d^4x\; M_1^2a^3\left[-r \left(2\tilde{Z}+Z\frac{d\ln M_1^2}{d\ln a}\right)\left( 3\Psi_2\Psi_1-a^2\Psi_2\partial^2 E_1-b^2\Psi_1\partial^2E_2 \right) \right. \nonumber\\
& \left.- Z\frac{\bar{N}}{(\bar{N}+r)}\partial_iB_2\partial^iB_1 -Z\frac{(\bar{N}-r)}{(H-H_b)} \left(H\Phi_1\left(3\Psi_2-b^2\partial^2 E_2\right) \right.\right.\nonumber\\
& \left. \left. +H_b\Phi_2\left(3\Psi_1-a^2\partial^2E_1\right)\right) - 2a^2b^2H^2\alpha_{L}\partial^2 E_1\partial^2 E_2 \right]. 
\end{align}
Here, we have two mass scales for each metric $M_1^2$ and $M_2^2$, and we have introduced the mass ratio $\nu^2=M_1^2/M_2^2$. In addition, for ease of comparison with massive bigravity, we have introduced two functions $Z$ and $\tilde{Z}$ such that: 
\begin{align}
& M_1^2\left(\bar{N}-r\right)Z= \hat{L}_{1KK\times},\label{DefZ}\\
& rM_1^2\left(Z\frac{1}{2}\frac{d\ln M_1^2}{d\ln a}+2\tilde{Z}\right)= \frac{1}{(H_b-H)} \left(3H+\frac{H_NH_b}{(H-H_b)}\right)\hat{L}_{1KK\times}+\frac{\dot{\hat{L}}_{1KK\times}}{(H_b-H)}\label{DefTildeZ},
\end{align}
where 
\begin{equation}
\hat{L}_{1KK\times}= \rho_m+P_m +2\dot{H}M_1^2.
\end{equation}
We omit the expression for $S_{\varphi}^{(2)}$ as it can be straightforwardly obtained from eq.~(\ref{SecondSmv2}). We emphasise, even though it may not be obvious, that these actions do depend on $2+3$ free independent functions. There is an explicit dependence on five parameters $M_A$, $Z$, $\tilde{Z}$ and $\alpha_L$, in addition to the four background functions $a$, $b$, $\bar{N}$ and $\bar{\varphi}$. However, $Z$ and $\tilde{Z}$ are dependent functions according to eq.~(\ref{DefZ})-(\ref{DefTildeZ}), one background function is dependent through eq.~(\ref{N4}), and $M_2$ is also dependent through one of the relations shown in Appendix \ref{App:ParamNonDer}, which is necessary to avoid derivative interactions. This relation is the following:
\begin{equation}\label{L1L2relation}
M_2^2=-\frac{\bar{N}}{2r^3} \frac{\left(\rho_m+P_m +2\dot{H}M_1^2 \right)}{\left(\dot{H}_b-H_NH_b\right)}.
\end{equation} 
From the parametrised action shown here we can see that the fields $\Phi_{A}$ and $B_{A}$ appear as auxiliary variables, whereas the fields $\Psi_{A}$ and $E_{A}$ have dynamical terms. A naive counting of DoFs might lead to a total of two physical propagating fields, but as we have mentioned before, this action propagates only one. This is because this action is such that after integrating out the four auxiliary fields $\Phi_{A}$ and $B_{A}$, the kinetic terms for $\Psi_2$ vanish and thus $\Psi_2$ becomes an auxiliary field. If we also integrated $\Psi_2$ out, the resulting action would have three dynamical fields $E_{A}$ and $\Psi_1$, from which two are gauge freedoms and one is a physical propagating DoF.

In addition, from eq.~(\ref{S1NonDer})-(\ref{S2NonDer}) we can see that all the kinetic terms have exactly the same structure as linearised Einstein-Hilbert. In fact, the two first lines in both equations correspond to the terms coming from two copies of linearised GR with generalised (time-dependent) Planck masses $M_A$, in addition to the self-interaction metric terms for $g_{1\mu\nu}$ from the matter action of eq.~(\ref{SecondSmv2}). The rest of the two lines in both equations represent then modifications to GR, which depend on the coupling parameters $Z$, $\tilde{Z}$ and $\alpha_{L}$. This last point is clear from the fact that all the interactions terms in eq.~(\ref{ST1T2NonDer}) depend on these three parameters. For this reason, an equivalent yet more intuitive parametrisation would be if we considered the five parameters $M_A$, $Z$, $\tilde{Z}$ and $\alpha_L$ to be independent, and all the background functions to be dependent.

This final action is a generalisation of massive bigravity \cite{deRham:2010ik, deRham:2010kj, Hassan:2011zd}, a bimetric theory propagating one massless graviton and one massive graviton. This theory propagates only one scalar field: the helicity-0 mode of the massive graviton. We recover the quadratic action and background of massive bigravity when the three interaction parameters take the following form: 
\begin{align}
&\alpha_{L}=0, \\
& Z= m^2\left(\beta_1+2\beta_2r+\beta_3r^2\right), \\
& \tilde{Z}= m^2\left(\beta_1+\beta_2\left(r+\bar{N}\right)+\beta_3r\bar{N}\right),
\end{align}
and when $M_1$ and $M_2$ are non-zero constants. Here, $\beta_{1,2,3}$ are dimensionless constants determining the coupling between the two metrics in the dRGT potential \cite{deRham:2010kj}, and $m$ is a constant mass scale degenerate with the parameters $\beta$s. Notice that we have given five constraints to recover massive bigravity, which fix completely the five free functions of time of the general bimetric action. We can also recover massive gravity (with only one dynamical metric) by setting $M_2=0$. We note, however, that such model has been ruled out due to the presence of an instability in which the helicity-0 mode of the massive graviton behaves as a ghost (i.e.~has a negative kinetic term) \cite{Fasiello:2012rw,Higuchi:1986py} when the non-dynamical metric is either FRW or de-Sitter. Furthermore, if the reference metric is Minkowski, massive gravity does not even allow a patially flat FRW solutions \cite{D'Amico:2011jj} (although a generalisation of massive gravity has been found to lead to viable cosmological solutions \cite{deRham:2014gla}.) 

We remark that the dRGT potential of massive gravity depends on two other constants, namely $\beta_0$ and $\beta_4$, which act as cosmological constants for both metrics. In our results we do not find the presence of such terms explicitly, but instead they appear as integration constants of equations (\ref{DefZ}) and (\ref{L1L2relation}), that give the time derivatives of the Friedmann equations of both metrics.

We notice that the parameter $\alpha_{L}$ is not present in massive bigravity, which means that it represents the linear term of a theory that either might propagate a Boulware-Deser ghost at the non-linear level, or that it is not fully diffeomorphism invariant, or that it propagates different DoFs. An example of the last case is the bimetric theory Eddington-inspired Born Infeld (EiBI) \cite{Vollick:2003qp, Banados:2010ix}. This is a bimetric theory for a massless graviton, and thus it does not propagate any scalar DoF, but it does introduce relevant modifications to GR -- specifically, in the strong-field regime. We can recover the EiBI quadratic action and background \cite{PhysRevD.89.024034} by setting:
\begin{align}
M_1^2\alpha_{L}&=\frac{r\bar{N}}{2H^2\kappa}, \label{EiBIParamsAL}\\
 \alpha_{E}&=\frac{1}{\kappa r^2M_2^2},\label{EiBIParamsAE}\\ 
M_1^2Z&=-\frac{1}{\kappa}\frac{r}{\bar{N}}\left(r+\bar{N}\right),\\
 M_1^2\tilde{Z}&= -\frac{1}{2r\kappa}\left(r\bar{N}+\frac{H_N^2}{2\bar{N}(H_b-H)^2}\right),
\end{align}
and when $M_1=0$ and $M_2$ is the constant Planck mass. Here, $\kappa$ is the coupling constant between the two metrics. From these equations we can see that in this theory there are no kinetic terms for the metric $g_{1\mu\nu}$, but there are non-derivative interactions terms. We clarify that the action for EiBI theory does not satisfy the extra constraint in eq.~(\ref{ConstraintOneDoF}), but instead $\alpha_{E}$ takes the value shown in eq.~(\ref{EiBIParamsAE}). In fact, in the EiBI action, the fields $\Psi_2$ and $E_2$ are the only dynamical variables, but they can both be fixed by the gauge freedom, leading then to an action with no scalar field propagating. Notice that since EiBI does not satisfy eq.~(\ref{ConstraintOneDoF}), it does not fall within the action presented in this section. Instead, EiBI is a specific case of the action in Appendix \ref{App:ParamNonDer} with $M_1=0$. Such an action depends on five free independent parameters, namely $M_2$, $Z$, $\tilde{Z}$, $\alpha_L$ and $\alpha_E$, or equivalently, $\alpha_L$, $\alpha_E$ plus three free independent background functions.

So far we have focused only on scalar perturbations, however, as we mentioned before, the constraints we have found on the parameters are also valid for vector and tensor perturbations. Thus, from our results it is possible to see that some of the three parameters modifying GR will also affect the vector and tensor perturbations. In particular, from the resulting action we can identify some relevant parameters of eq.~(\ref{LT1T20}) and see that vector perturbations are coupled through $P_{S_2S_1}$ whereas tensor perturbations have a coupling through $P_{h_1h_2\times}$. Specifically, for the subclass of theories addressed in this section, these two parameters take the following form:
\begin{align}
P_{S_2S_1}&= -a^3M_1^2\frac{Z\bar{N}r^2}{(\bar{N}+r)},\label{VectorCoupling}\\
P_{h_1h_2\times}&=\frac{M_1^2a^3}{8r}\left(2\tilde{Z}+Z\frac{d\ln M_1^2}{d\ln a}\right)-\frac{3}{4r^2}M_1^2a^3H^2\alpha_{L}.\label{TensorPertCoeff}
\end{align}
From this we conclude that $Z$ generates interactions between vector perturbations, while the three parameters $Z$, $\tilde{Z}$ and $\alpha_L$ generate interactions between tensor perturbations. This result is consistent with previous studies in massive bigravity, and its two branches of solutions. In the so-called branch I, where $Z=0$, it has been found that scalar and vector perturbations behave in the same way as in GR, while tensor perturbations are coupled and evolve in a different way \cite{vonStrauss:2011mq, Cusin:2015tmf}. This is in fact what we find from our results: vector perturbations are not coupled if $Z=0$ because of eq.~(\ref{VectorCoupling}). While scalar perturbations would have a coupling with $\tilde{Z}$, in this case that coupling happens to be irrelevant after integrating out the auxiliary fields, and thus scalars behave as in GR when $Z=0$. Finally, tensor perturbations do have a non-trivial coupling with $\tilde{Z}$, which indeed affects their evolution. In the so-called branch II, where $Z\not=0$, the three types of perturbations are coupled and differ from GR \cite{Lagos:2014lca, Cusin:2014psa}. 

Finally, we emphasise that even though a detailed analysis in vector and tensor perturbations is necessary, scalar perturbations carry crucial information. From eq.~(\ref{VectorCoupling})-(\ref{TensorPertCoeff}) we can see that by observing vector perturbations, we can analyse the behaviour of the parameter $Z$, while from tensor perturbations we cannot discriminate between $\tilde{Z}$ and $\alpha_L$, as they appear in the same interaction term. This suggests that an observational test to discriminate if $\alpha_L$ is present or not can be done by analysing scalar perturbations alone. 


\section{Discussion}
\label{sec:discussion}
In this paper we applied the method developed in \cite{Lagos:2016wyv} to bimetric theories. We calculated the most general diffeomorphism-invariant quadratic action with two metrics, around a homogeneous and isotropic background, and leading to, up to, second-order equations of motion. For simplicity, we assumed that only one of the metrics was coupled to matter, a minimally coupled scalar field, although generalisations to perfect fluids or double couplings should be straightforward. Following the standard SVT decomposition for cosmological perturbations, we focused on scalar perturbations, and found that the final action depends on 29 free parameters (functions of time), in addition to three parameters determining the background evolution, and propagates at most four scalar physical DoFs. Due to the no-go theorem for healthy Lorentz-invariant bimetric theories with derivative interactions, in this paper we focused on the subclass of bimetric theories without derivative interactions. In this case, we find that the number of free parameters in the quadratic action greatly reduces from 29 to 3, namely $M_1$, $\alpha_L$ and $\alpha_E$, in addition to three extra free parameters that determine the evolution of the background. The resulting action propagates at most two scalar DoFs, which suggests the presence of an unstable mode due to the fact that all well-known bigravity theories propagate at most one scalar DoF. For this reason, we focused on subclasses of theories that propagate one or no scalar field. 

In order to construct actions with only one propagating DoF, we imposed an extra constraint on the free parameters, which fixed the value of $\alpha_E$. In this case, the most general action has only two free parameters, and we found that it is a generalisation of the quadratic action of massive bigravity. We recovered massive bigravity when $\alpha_L=0$ and $M_1$ is a constant mass scale. We found that the presence of the parameter $\alpha_L$ affects the evolution of scalar and tensor perturbations, and even though it is not present in massive bigravity, it is present in other bimetric models such as EiBI theory. We also looked at cases in which the bimetric action propagates no scalar field. We found that when $M_1=0$, all the kinetic terms of one of the metrics vanished and, as result, the gravitational action does not propagate any physical scalar DoF. Such an action depends on two free parameters, and represents a generalisation of the EiBI theory. 


\begin{table}[h!]
\centering
\begin{tabular}{| c || c || c |}
\hline
Fields & Free Functions & Theory \\ \hline \hline
 $g_{\mu\nu}$, $\chi$ & $M$, $\alpha_{\{K, T, B\}}+2$ & Horndeski \\ \hline
$g_{\mu\nu}$, $A^{\mu}$ & $M$, $\alpha_{\{ T, H\}}$, $\alpha_{D3}$, $\alpha_{V{\{0, 1,2,3\}}}+2$ & Generalised Proca\\ \hline
$g_{\mu\nu}$, $A^{\mu}$, $\lambda$ & $M$, $\alpha_{V{\{3,4,5\}}}+1$ & Einstein-Aether\\ \hline
\rowcolor[gray]{.95}
$g_{1\mu\nu}$, $g_{2\mu\nu}$ & $M_1$, $\alpha_{L}+3$ & Massive bigravity\\ \hline
\rowcolor[gray]{.95}
$g_{1\mu\nu}$, $g_{2\mu\nu}$ & $\alpha_E$, $\alpha_{L}+3$ & EiBI\\ \hline
\end{tabular}
\caption{\label{SummaryTable} In this table we compile parametrised models for scalar-tensor, vector-tensor and bimetric theories of gravity, that are invariant under linear coordinate transformations, lead to second-order derivative equations, and propagate at most one scalar DoF. The first column indicates the field content of the gravitational theory. The second column shows the free functions parametrising the quadratic action for cosmological perturbations, while $+1$, $+2$ or $+3$ counts the number of extra free functions determining the background (and in turn affecting the perturbations). In most cases there is one free mass parameter $M$ or $M_1$, and extra adimensional parameters that we term $\alpha$. The third column shows examples of non-linear completions that are encompassed by the corresponding parametrisation. While the first three rows show results obtained in \cite{Lagos:2016wyv}, the last grey rows show the results found in this paper. We emphasise that in the vector-tensor case we dot not assume a dependence on the parameters $\alpha_{D1}$ and $\alpha_{D2}$ (see \cite{Lagos:2016wyv}), as we will set them to zero in order to get an action propagating one scalar DoF with the same structure as Generalised Proca.}
\end{table}

Combining the results from this paper with those of \cite{Lagos:2016wyv} we have been able to extend the widely-used parametrisation of \cite{Gleyzes:2014rba} originally proposed for Horndeski theories. It is now possible to construct a complete action for linear perturbations for general gravity with one propagating degree of freedom arising from either a scalar, vector or tensor field as can be seen in Table \ref{SummaryTable}. It has been shown that adding more interacting tensor fields will necessarily lead to more propagating scalar DoFs \cite{Noller:2013yja}. We expect the same to be true when adding vector fields, unless further gauge symmetries arise, such as invariance under U(1) transformations. Similarly, the addition of more scalar fields should lead to the propagation of extra DoFs, unless they appear as auxiliary variables. Therefore, even though there may not exist a theorem that would prevent adding more fields to the actions we have studied, while maintaining only one propagating DoF and its diffeomorphism invariance, we have been unable to find non-trivial such examples. Hence there is a possibility that our combined parametrisation for theories with one propagating degree of freedom is complete. 

Our action should allow us to identify the subspace of effective parameters in the Parametrised Post-Friedman (PPF) approach \cite{Baker:2011jy,Baker:2012zs} which is, at the moment, still the most general parametrisation of gravitational theories currently available. Ultimately it should be possible develop a numerical tool, along the lines of EFTCAMB \cite{Hu:2013twa} or HiCLASS \cite{Zumalacarregui:2016pph}, which can be used for analysing data from future large-scale structure surveys such as Euclid, SKA, LSST and WFIRST, allowing us to test and compare the performance of these theories. 

Special care must be taken when constraining modified gravity theories with cosmological data in the context of linear perturbation theory as non-linear effects will become relevant on small to intermediate scales. In fact, in modified gravity theories non linearities can become relevant at much larger scales than those in the standard $\Lambda$CDM model; the linear approximation of a modified gravity theory can give an inaccurate prediction of the universe even at scales as large as $k\sim 0.05h$/Mpc at present (see \cite{2012JCAP...10..002B, 2013JCAP...04..029B}). This happens because many modified gravity theories propagate more DoFs than GR and these extra DoFs undergo a non-linear process known as screening (such as the Chameleon mechanism in scalar-tensor theories or the Vainshtein mechanism in bimetric theories), which, depending on the specific theory, can have a substantial effect in regimes which seem, a priori, linear. While we emphasise that the tools presented in this paper can be used to predict the evolution of perturbations and thus constrain modified gravity theories at sufficiently large scales, a more detailed and accurate understanding of the effects of screening (such as in \cite{Schmidt:2009sg, Barreira:2013eea, Li:2013tda, Falck:2014jwa, Winther:2014cia}) must also be used in order to improve and extend these results (see \cite{Alonso:2016suf} for an attempt at including these effects yet using linear perturbations for scalar-tensor theories).

\begin{acknowledgments}
We are extremely grateful for conversations with Hans Winther, Tessa Baker, Johannes Noller and Jerome Gleyzes. ML was funded by Becas Chile, CONICYT. PGF acknowledges support from Leverhulme, STFC, BIPAC and the ERC.

\end{acknowledgments}

\appendix
\section{Scalar perturbations}
\label{app:metricperts}
In this section we show relevant quantities in terms of the four linear scalar perturbations of each metric. Following the standard SVT decomposition, we consider linear perturbations around a homogeneous and isotropic background, and write the metrics in the following way:

\begin{align}\label{Def4Pert}
&ds_1^2=-\left(1+\Phi_1\right)dt^2+2\partial_i B_1 dt dx^i +a^2\left[\left(1-2\Psi_1\right)\delta_{ij}+2\partial_i\partial_j E_1\right]dx^i dx^j,\\
&ds_2^2=-\bar{N}^2\left(1+\Phi_2\right)dt^2+2\bar{N}\partial_i B_2 dt dx^i +b^2\left[\left(1-2\Psi_2\right)\delta_{ij}+2\partial_i\partial_j E_2\right]dx^i dx^j,
\end{align}
where $a$, $b$ and $\bar{N}$ are background quantities and depend only on the time $t$, whereas the 8 perturbations $\Phi_A$, $B_A$, $\Psi_A$ and $E_A$ (for $A=\{1,2\}$) depend on time and space. From the ADM decomposition of eq.~(\ref{ADMDecomposition}), we can find the results of eq.~(\ref{ADMmetric1})-(\ref{ADMmetric2}), and also express all the relevant quantities used throughout the paper in terms of the scalar metric perturbations. Here we give a list of quantities, that appear in the quadratic gravitational action $S_G^{(2)}$, in terms of the scalar perturbations:
 \begin{eqnarray}
 \delta_2 N_1 &= &-\frac{1}{2}\left(\Phi_1^2+\bar{h}^{ij}_1\partial_iB_1\partial_jB_1\right), \nonumber\\
 \delta \sqrt{h_1}&=&a^3\left[-3\Psi_1- a^2\partial^2 E_1\right],\nonumber\\
 \delta_2 \sqrt{h_1} & =& a^3\left[ \frac{3}{2}\Psi_1^2 - \frac{1}{2}a^4(\partial^2 E_1)(\partial^2 E_1)-a^2\Psi_1 \partial^2 E_1\right],\nonumber\\
 \delta K^i_{\phantom{i}1j}&=&-({\dot \Psi_1}+H\Phi_1)\delta^i_{\phantom{i}j} + a^2\bar{h}^{il}_1\partial_l\partial_j{\dot E_1}-\bar{h}^{il}_1\partial_l\partial_j B_1,\nonumber \\
 \delta K_1&=& -3({\dot \Psi_1}+H\Phi_1) + a^2 \partial^2 {\dot E_1}-\partial^2 B_1, 
 \end{eqnarray}
 \begin{eqnarray}
 \delta_2 N_2 &=& -\frac{\bar{N}}{2}\left(\Phi_2^2+\bar{h}^{ij}_2\partial_iB_2\partial_jB_2\right), \nonumber\\
 \delta \sqrt{h_2}&=&b^3\left[-3\Psi_2- b^2\partial^2 E_2\right], \nonumber\\
 \delta_2 \sqrt{h_2} & =& b^3\left[ \frac{3}{2}\Psi_2^2 - \frac{1}{2}b^4(\partial^2 E_2)(\partial^2 E_2)-b^2\Psi_2 \partial^2 E_2\right], \nonumber\\
 \delta K^i_{2\phantom{i}j}&=&-\frac{1}{\bar{N}}\left({\dot \Psi_2}+H_b\Phi_2\right)\delta^i_{\phantom{i}j} + \frac{b^2}{\bar{N}}\bar{h}^{il}_2\partial_l\partial_j{\dot E_2}-\bar{h}^{il}_2\partial_l\partial_j B_2,\nonumber \\
 \delta K_2&=& -\frac{3}{\bar{N}}({\dot \Psi_2}+H_b\Phi_2) +\frac{b^2}{\bar{N}} \partial^2 {\dot E_2}-\partial^2 B_2, 
 \end{eqnarray}
where $\bar{h}^{ij}_1=\frac{\delta^{ij}}{a^2}$ and $\bar{h}^{ij}_2=\frac{\delta^{ij}}{b^2}$ represent the background spatial metrics, and $\partial^2=\partial^i\partial_i$, where the indices are lowered and raised using the background metric of the corresponding field the derivative is acting on. Also, a single $\delta$ stands for linear perturbations, while $\delta_2$ stands for quadratic perturbations.

\section{Dictionary of parameters}\label{App:parameters}
In this section we give expressions for the 29 parameters the final bimetric action depends on. The names we have given to those parameters are the following:
\begin{equation}
M_A^2, \; \alpha_{H_A}, \; \alpha_{T_A}, \; \alpha_{B_A}, \; \alpha_{K_A}, \; \alpha_{L}, \; \alpha_{E},\; \alpha_{i,A}, \; \alpha_{j},
\end{equation}
where $A=\{1,2\}$, $i=\{1..4\}$ and $j=\{5..13\}$. In terms of the original coefficients $L$s and $P$s these parameters can be expressed as:
\begin{align}
 M_A^2& = 2L_{AKK\times}, \\
 \alpha_{H_A}&=-\frac{1}{M_A^2}\left(2L_{Ah\partial^2N\times}+3L_{Ah\partial^2N+}\right)-1,\\
 \alpha_{T_A}&=-\frac{4}{M_A^2}\left(L_{Ah\partial^2h\odot}+3L_{Ah\partial^2h\times}+9L_{Ah\partial^2h+}\right)-1,\\
 \alpha_{B_A}&=\frac{1}{2H_AM_A^2}\left[H_{AN}L_{AK\dot{N}}+T_{ANK}-\frac{H_A}{\bar{N}_A^2}(2L_{AKK\times}+3L_{AKK+})\right]-1,\\
\alpha_{K_A}& =\frac{1}{M_A^2H_A^2\bar{N}_A^2}\left[\bar{N}_A^4L_{ANN}-9H_A^2\left(L_{AKK+}+2L_{AKK\times}\right)\right.\nonumber\\
& \left. -12H_A^2\left(\bar{N}_A^2-1\right)L_{AKK\times}+\bar{N}^3_AH_{AN}L_{A\dot{N}\dot{N}}\right],\\
\alpha_{L}&=-\frac{1}{2M_1^2H^2}\left(\bar{L}_1-4L_{1hh+}-8L_{1hh\times}\right), \\
\alpha_{E}&=-\frac{2}{M_2^2}\left[ \bar{N}^2T_{2Nh}-H_b\left(2L_{2hK\times}+3L_{2hK+}\right)\right], \\ 
 \alpha_{1,A}&=-\frac{1}{M_A^2}L_{AK\dot{N}},\\
 \alpha_{2, A}&=-\frac{1}{M_A^2}\left(L_{Ah\partial^2h\odot}+2L_{Ah\partial^2h\times}+3L_{Ah\partial^2h+}\right), \\
 \alpha_{3,A}&= \frac{\bar{N}_A}{M_A^2}\left(2L_{Ah\partial^2N\times}+L_{Ah\partial^2N+}\right),\\
 \alpha_{4,A}&= \frac{1}{M_A^2}L_{A\partial N\partial N},\\
\alpha_{5}&= \frac{1}{M_1^2}L_{1\dot{S}\dot{S}}= \frac{\bar{N}r^5}{M_1^2} L_{2\dot{S}\dot{S}}= -\frac{1}{M_1^2a^3} P_{\dot{S}_1\dot{S}_2}, \label{Bdot2rel}\\
 \alpha_{6}&= \frac{1}{M_1^2} L_{1\dot{N}\dot{N}}= \frac{\bar{N}^3r^3}{M_1^2} L_{2\dot{N}\dot{N}}= -\frac{\bar{N}}{M_1^2a^3} P_{\dot{N}_1\dot{N}_2},\label{Phidot2rel}\\
 \alpha_{7}&= -\frac{2}{M_1^2}\left(L_{1\partial \dot{S}h+}+2L_{1\partial \dot{S}h\times}\right)+\frac{1}{M_1^2H}\left(L_{1\partial S h+}+2L_{1\partial S h\times} -L_{1hK+}-2L_{1hK\times}\right), \\
 \alpha_{8}&=\frac{1}{M_1^2a^3}(r^2P_{h_2\partial^2N_1\times}+3P_{h_2\partial^2N_1+}),\\
 \alpha_{9}&= \frac{1}{M_1^2a^3}P_{h1\partial^2N2+},\\
 \alpha_{10}&= \frac{1}{M_1^2}L_{1h\partial^2h\times},\\
 \alpha_{11}&=\frac{1}{M_1^2a^3}P_{K_1K_2+} , \\
 \alpha_{12}&=\frac{1}{M_1^2a^3}\left(3P_{K_1K_2+}+P_{K_1K_2\times}\right),\\
 \alpha_{13}&=\frac{1}{M_1^2a^3}\left[r^6\left(P_{h_2\partial^2h_1\odot}+3P_{h_2\partial^2h_1\times 2}\right)+9r^4P_{h_2\partial^2h_1+}+3P_{h_2\partial^2h_1\times 1}\right],
\end{align}
where we have defined $\bar{N}_A$ such that $\bar{N}_1=1$ and $\bar{N}_2=\bar{N}$, thus $H_{1N}=0$, $H_{1N}=H_N$, and also $H_1=H$ and $H_2=H_b$. Here, we have also introduced the ratio of the scale factors $r=b/a$. The parameters $M_A$ have mass dimensions and appear multiplying the whole quadratic action, whereas all the parameters $\alpha$s are dimensionless and are the couplings coefficients of the different interactions terms for the fields. For instance, the parameters $\alpha_{H_A}$, $\alpha_{T_A}$, $\alpha_{B_A}$, and $\alpha_{K_A}$ determine the interactions terms given by $\Psi_A\partial^2\Phi_A$, $\Psi_A\partial^2\Psi_A$, $\Phi_A\dot{\Psi}_A$ and $\Phi_A^2$, respectively. These parameters are generalisations of those present in the parametrisation of dark energy models of \cite{Gleyzes:2014rba, Lagos:2016wyv}. 

Finally, we comment on the fact that these 29 parameters can have different expressions in terms of the functions $L$s and $P$s, if we use relations between them given by the Noether constraints. This is why parameters $\alpha_{5}$ and $\alpha_6$ have three equivalent expressions in equations (\ref{Bdot2rel}) and (\ref{Phidot2rel}). For these specific parameters, these three expressions show that the dynamical terms of the fields $B_1$ with $B_2$ and $\Phi_1$ with $\Phi_2$ are related to each other, signalling the fact that the dynamical terms of these fields appear in a specific combination in the action, and thus there is a field redefinition such as eq.~(\ref{Phi3B3}), that can make two fields appear as auxiliary fields (without time derivatives) instead of dynamical fields.

\section{Complete action without derivative interactions}\label{App:ParamNonDer}
If we avoid derivative interactions between both metrics, then 26 of the 29 previous parameters are fixed. Specifically, we find that the following 26 constraints:
\begin{align}
 M_2^2 &=-\frac{\bar{N}}{2r^3}\frac{\left(2\dot{H}M_1^2+\rho_m+P_m\right)}{\left(\dot{H}_b-H_NH_b\right)}, \label{M2M1rel}\\ 
 \alpha_{H_1} &= 0, \; \alpha_{H_2} = -\frac{(\bar{N}-1)}{\bar{N}} \label{AlphaHNonDer}, \\
 \alpha_{T_A} &= \frac{d\ln M_A^2}{d\ln a_A},\\
 \alpha_{B_1} &= 0, \; \alpha_{B_2} = -\frac{\left(\bar{N}-1\right)\left(\bar{N}+1\right)}{\bar{N}^2}, \\
 \alpha_{K_1} &= \frac{6H_b\dot{H}}{H_NH^2}+\frac{1}{2H^2}(2\rho_m+P_m)+\frac{3}{2}(\rho_m+P_m)\frac{H_b}{H_NH^2}-3\alpha_E\frac{(H-H_b)}{\nu^2\bar{N}H_NH^2},\\
 \alpha_{K_2} &= -\frac{6}{\bar{N}^2H_NH_b}\left[H_bH_N\left(\bar{N}^2-2\right)+\dot{H}_b \right]-3\alpha_E\frac{(H-H_b)}{\bar{N}H_NH_b^2},\\
 \alpha_{1,A} &= \alpha_{2,A} = \alpha_{3,A} = \alpha_{4,A}= 0,\\
 \alpha_{5} &=\alpha_{6} = \alpha_{7} = \alpha_{8} = \alpha_{9} = 0, \\
 \alpha_{10} &=\frac{1}{8}\left(\frac{d\ln M_1^2}{d\ln a}+1\right), \\
 \alpha_{11} &= \alpha_{12} = \alpha_{13}=0,
\end{align}
where we have defined the scale factor $a_A$ such that $a_1=a$ and $a_2=b$, and we have introduced the ratio of the mass scales $\nu^2=M_1^2/M_2^2$. We can see that the quadratic bimetric action without derivative interactions depends only on three independent free parameters: $\alpha_{E}$, $\alpha_{L}$ and $M_1$, in addition to the four background functions $a$, $b$, $\bar{N}$ and $\bar{\varphi}$, which give three additional independent free functions, due to the background equation (\ref{N4}).

The resulting quadratic action can be written as:
\begin{equation}\label{App:TotalActionNonDer}
S^{(2)}=S_{T_1}^{(2)}+ S_{T_2}^{(2)}+ S_{T_1T_2}^{(2)}+ S_{\varphi}^{(2)},
\end{equation}
where $S_{T_A}^{(2)}$ is the action for the self-interaction terms of the metric $g_{A\mu\nu}$, whereas $S_{T_1T_2}^{(2)}$ is the action for the interaction terms between both metrics, and $S_{\varphi}^{(2)}$ includes all the terms involving the matter perturbation $\delta \varphi$ in eq.~(\ref{SecondSmv2}). Notice that $S_{T_1}^{(2)}$ does include the quadratic terms of the metric perturbations coming from the matter action $S_m^{(2)}$. These actions are given by:
\begin{align}\label{App:S1NonDer}
S_{T_1}^{(2)}&=\int d^4x\; a^3M_1^2\left[-3\dot{\Psi}_1^2-6H\dot{\Psi}_1\Phi_1+2a^2\partial^2\dot{E}_1(\dot{\Psi}_1+H\Phi_1)-2\dot{\Psi}_1 \partial^2B_1 \right. \nonumber\\
& -\left(1+\frac{d\ln M_1^2}{d\ln a}\right)\Psi_1\partial^2 \Psi_1 +2\Phi_1\partial^2\Psi_1-2H\Phi_1\partial^2 B_1- \left(3H^2 - \frac{\dot{\bar{\varphi}}^2}{2M_1^2} \right)\Phi_1^2\nonumber\\
& + \frac{r^2Z}{2(\bar{N}+r)}\left(\partial^i B_1\right)\left(\partial_iB_1\right) + r \left(Z\frac{d\ln M_1^2}{d\ln a}+2\tilde{Z}\right) \Psi_1\left(\frac{3}{2}\Psi_1-a^2\partial^2E_1\right) \nonumber\\
& +\frac{(\bar{N}-r)}{H-H_b}HZ\Phi_1\left(3\Psi_1- a^2\partial^2E_1\right) +a^4\alpha_{L}H^2\left(\partial^2E_1\right)^2 \nonumber\\
& +\left(H_b\left(\bar{N}-r\right)Z-\frac{\left(H-H_b\right)}{\nu^2}r^3\alpha_E\right)\left(\frac{3}{2}\frac{1}{H_N}\Phi_1^2+\frac{H_N}{(H-H_b)^2}\Psi_1\left(\frac{3}{2}\Psi_1-a^2\partial^2E_1\right)\right. \nonumber\\
& \left. \left. -\frac{1}{(H-H_b)}\Phi_1\left(3\Psi_1- a^2\partial^2E_1\right) \right)\right], 
\end{align}
\begin{align}\label{App:S2NonDer}
S_{T_2}^{(2)}&=\int d^4x\; \bar{N}b^3M_2^2\left[-3\frac{\dot{\Psi}_2^2}{\bar{N}^2}-6\frac{H_b}{\bar{N}^2}\dot{\Psi}_2\Phi_2+2\frac{b^2}{\bar{N}^2}\partial^2\dot{E}_2(\dot{\Psi}_2+H_b\Phi_2)-2\frac{\dot{\Psi}_2}{\bar{N}} \partial^2B_2 \right. \nonumber\\
& -\left(1+\frac{d\ln M_2^2}{d\ln b}\right)\Psi_2\partial^2 \Psi_2 +2\Phi_2\partial^2\Psi_2-2\frac{H_b}{\bar{N}}\Phi_2\partial^2 B_2-3\frac{H_b^2}{\bar{N}^2}\Phi_2^2 \nonumber\\
& + \frac{\nu^2\bar{N}Z}{2r^3(\bar{N}+r)}\left(\partial^i B_2\right)\left(\partial_iB_2\right)+ \frac{\nu^2}{\bar{N}r^2}\left(Z\frac{d\ln M_1^2}{d\ln a}+2\tilde{Z}\right) \Psi_2\left(\frac{3}{2}\Psi_2-b^2\partial^2E_2\right) \nonumber\\
& + H_b\nu^2Z\frac{(\bar{N}-r)}{\bar{N}r^3(H-H_b)}\Phi_2\left(3\Psi_2- b^2\partial^2E_2\right) +\frac{r}{\bar{N}}\alpha_{L}a^4H^2\nu^2\left(\partial^2E_2\right)^2 \nonumber\\
& +\left(H_b\left(\bar{N}-r\right)Z-\frac{\left(H-H_b\right)}{\nu^2}r^3\alpha_E\right)\left(\frac{3}{2}\frac{1}{H_N}\Phi_2^2+\frac{H_N}{(H-H_b)^2}\Psi_2\left(\frac{3}{2}\Psi_2-b^2\partial^2E_2\right)\right. \nonumber\\
& \left. \left. -\frac{1}{(H-H_b)}\Phi_2\left(3\Psi_2- b^2\partial^2E_2\right) \right)\frac{1}{r^3\nu^2\bar{N}} \right], 
\end{align}
\begin{align}\label{App:ST1T2NonDer}
S_{T_1T_2}^{(2)}&=\int d^4x\; M_1^2a^3\left[-r \left(2\tilde{Z}+Z\frac{d\ln M_1^2}{d\ln a}\right)\left( 3\Psi_2\Psi_1-a^2\Psi_2\partial^2 E_1-b^2\Psi_1\partial^2E_2 \right) \right. \nonumber\\
& - Z\frac{\bar{N}}{(\bar{N}+r)}\partial_iB_2\partial^iB_1 -Z\frac{(\bar{N}-r)}{(H-H_b)} \left(H\Phi_1\left(3\Psi_2-b^2\partial^2 E_2\right) \right.\nonumber\\
& \left. +H_b\Phi_2\left(3\Psi_1-a^2\partial^2E_1\right)\right) - 2a^2b^2H^2\alpha_{L}\partial^2 E_1\partial^2 E_2 \nonumber\\
& + \left(H_b\left(\bar{N}-r\right)Z-\frac{\left(H-H_b\right)}{\nu^2}r^3\alpha_E\right)\left( \frac{3}{H_N}\Phi_1\Phi_2+ \left(H_b-H \right)\left(\Phi_1\left(3\Psi_2-b^2\partial^2E_2\right)\right.\right. \nonumber\\
& \left. \left. \left. + \Phi_2\left( 3\Psi_1-a^2\partial^2E_1\right)\right) +H_N\left( 3\Psi_2\Psi_1-a^2\Psi_2\partial^2 E_1-b^2\Psi_1\partial^2E_2\right)\right)\frac{1}{\left(H_b-H\right)^2}\right]. 
\end{align}
Here, we have two mass scales for each metric $M_1^2$ and $M_2^2$, and we have introduced the mass ratio $\nu^2=M_1^2/M_2^2$, and the scale factor ratio $r=b/a$. In addition, for ease of comparison with massive bigravity, we have introduced two functions $Z$ and $\tilde{Z}$ such that: 
\begin{align}
& M_1^2\left(\bar{N}-r\right)Z= \hat{L}_{1KK\times},\label{App:DefZ}\\
& rM_1^2\left(Z\frac{1}{2}\frac{d\ln M_1^2}{d\ln a}+2\tilde{Z}\right)= \frac{1}{(H_b-H)} \left(3H+\frac{H_NH_b}{(H-H_b)}\right)\hat{L}_{1KK\times}+\frac{\dot{\hat{L}}_{1KK\times}}{(H_b-H)}\label{App:DefTildeZ},
\end{align}
where 
\begin{equation}
\hat{L}_{1KK\times}= \rho_m + P_m +2\dot{H}M_1^2.
\end{equation}
We omit the expression for $S_{\varphi}^{(2)}$ as it can be straightforwardly obtained from eq.~(\ref{SecondSmv2}). We emphasise that the total parametrised action depends on $3+3$ free independent functions of time. There is a dependence on six parameters $M_A$, $Z$, $\tilde{Z}$, $\alpha_L$ and $\alpha_E$, in addition to the four background functions $a$, $b$, $\bar{N}$ and $\bar{\varphi}$. However, $Z$ and $\tilde{Z}$ are dependent functions according to eq.~(\ref{App:DefZ})-(\ref{App:DefTildeZ}), one background function is dependent through eq.~(\ref{N4}), and $M_2$ is also dependent through eq.~(\ref{M2M1rel}). Equivalently, we can consider the six independent parameters to be $M_A$, $Z$, $\tilde{Z}$, $\alpha_L$ and $\alpha_E$, while the background functions would be dependent. 

From the parametrised action shown here we can see that the fields $\Phi_{A}$ and $B_{A}$ appear as auxiliary variables, whereas the fields $\Psi_{A}$ and $E_{A}$ have dynamical terms. This means that, in general, this action propagates two physical scalar fields. Nevertheless, there are some trivial cases in which no scalar is propagated. This happens if either $M_1$ or $M_2$ vanishes, and thus all the dynamical terms of one of the metrics vanish (although these metrics can still have non-derivative terms as long as the quantities $ZM_A^2$, $\tilde{Z}M_A^2$, $\alpha_LM_A^2$ or $\alpha_EM_A^2$ are finite). As it is shown in Section \ref{sec:non-der}, Eddington-inspired Born Infeld theory is an example of a gravity model with $M_1=0$ that does not propagate any scalar DoF. On the other hand, there are also cases that propagate only one scalar DoF, such as massive bigravity, which is presented in Section \ref{sec:non-der}.

 
\section{The effect of derivative interactions}
\label{sec:der}
As it is shown in Section \ref{sec:non-der}, when there are no derivative interactions between the metrics, the kinetic terms of each metric correspond to linearized Einstein-Hilbert with a generalised Planck mass. In the context of massive gravity, work towards theories that go beyond the ordinary Einstein-Hilbert terms include \cite{Folkerts:2011ev, Hinterbichler:2013eza, Kimura:2013ika}, although later on in \cite{deRham:2013tfa} it was shown that it is not possible to non-linearly complete the specific terms studied in previous analyses without reintroducing the Boulware-Deser ghost below the cutoff-scale of the effective field theory, concluding that there can be no new healthy Lorentz-invariant derivative interactions in the metric formulation. 

For completeness, in this section we briefly discuss the case in which we do allow derivative interactions, which might be relevant in the context of Lorentz-breaking theories (although keeping linearised diffeomorphism invariance) due to the no-go theorem of \cite{deRham:2013tfa}. Lorentz-breaking massive gravity in flat space has been studied before as it avoids the VDVz discontinuity and improves the strong coupling scale of the effective field theory \cite{Rubakov:2004eb,Dubovsky:2004sg,Rubakov:2008nh}, giving an interesting alternative to the standard Fierz-Pauli theory of massive gravity. Possible generalisations to curved space have also been studied \cite{Blas:2009my}, but they break linear diffeomorphism invariance, and thus they would not be included in the class of theories studied in this paper. 

In what follows, we will study theories with derivative interactions that propagate only one scalar DoF. Such actions can be constructed in different ways, but as an example we mention a case that has a similar structure to massive bigravity, that is, where the fields $\Phi_{A}$ and $B_{A}$ are auxiliary variables. For this to happen the 29 parameters presented on Appendix \ref{App:parameters} must satisfy the following constraints:
\begin{equation}
\alpha_5=\alpha_6=0,\; \alpha_{1,A}=0, \; \alpha_{4,1}\nu^2=r\bar{N}^3\alpha_{4,2}.
\end{equation}
Thus, the most general action satisfying these constraints will depend on 24 free parameters, in addition the background free functions, and will propagate at most two scalar DoFs. In such actions the fields $E_A$ and $\Psi_A$ will appear as dynamical fields. In order to construct ghost-free actions propagating only one scalar DoF we impose that one of the dynamical fields is an auxiliary variable after integrating out the four auxiliary fields $B_A$ and $\Phi_A$. Again, following the structure of massive bigravity, we impose that $\Psi_2$ is an auxiliary variable. We find that this can happen when different sets of constraints for the parameters are satisfied. For instance, this happens if:
\begin{align}\label{DerivativeCoeffsExample}
\alpha_{4,1}&=\alpha_{12}=0, \nonumber\\
\alpha_{K_2}H_NH_b^2&=3\left[2H_b^2H_N(2\alpha_{B_2}+1)-2H_b\dot{H}_b(\alpha_{B_2}+1)-\alpha_{E}(H-H_b) \right],\nonumber\\
\alpha_{3,2}r^3H&=H_b\bar{N}\frac{(6\alpha_{B_2}-\alpha_{K_2})(\alpha_{H_2}+1)}{6(\alpha_{B_2}+1)}+H_br^2\nu^2\alpha_8\frac{(\alpha_{K_2}-12\alpha_{B_2}-6)}{6\bar{N}(\alpha_{B_2}+1)}+2H\nu^2\alpha_9,\nonumber\\
 \alpha_{K_1}H^2M_1^2&=(2\rho_m+P_m)+\bar{N}r^3M_2^2\frac{(\alpha_{K_2}-6\alpha_{B_2})^2}{(\alpha_{K_2}-12\alpha_{B_2}-6)}+6H^2M_1^2\alpha_{B_1}.
\end{align}
Actions satisfying these five constraints propagate only one scalar DoF and, in general, have derivative interactions between both metrics. In addition, we notice that the values for the parameters in Appendix \ref{App:ParamNonDer} set to avoid derivative interactions, along with the extra constraint in eq.~(\ref{ConstraintOneDoF}), are a particular case of the constraints presented here. Therefore, actions satisfying eq.~(\ref{DerivativeCoeffsExample}) are a direct generalisation of the non-derivative action shown in Section \ref{sec:non-der}. 

In order to illustrate the form that the action can take now, equation (\ref{ActionDerExample}) shows the extra interaction terms that appear in the quadratic action, compared to the terms in eq.~(\ref{TotalActionNonDer}), due to the new set of constraints given eq.~(\ref{DerivativeCoeffsExample}), when $\alpha_{9}$ is a non-zero constant (and thus $\alpha_{3,2}\not=0$ due to the constraints) and when the rest of the parameters take the same value as those of Appendix \ref{App:ParamNonDer}:
\begin{align}\label{ActionDerExample}
\Delta S^{(2)}_{T_1T_2}&=\int d^4x \; 16a^5\alpha_9M_1^2\left[\frac{\bar{N}}{\left(\bar{N}^2-r^2\right)} \left(\bar{N}\partial^2B_2\partial^2\dot{E}_1+ \partial^2B_1\partial^2\dot{E}_2\right) +H_b\partial^2B_2\partial^2E_1 \right. \nonumber\\
& + 3\frac{\bar{N}^2\left(H-H_b\right)}{\left(\bar{N}^2-r^2\right)a^2}\partial^2B_2\left(\Psi_1+\frac{Hr^2}{\left(\bar{N^2}-r^2\right)}B_1\right) -\bar{N}\partial^2\Phi_2\partial^2E_1 \nonumber\\
& \left. -\frac{a^2r^2}{\bar{N}}H_b\partial^2E_2\partial^2 \dot{E}_1 +\frac{a^2r^2}{\bar{N}}\left(\dot{H}_b-H_NH_b+3H_b^2\right)\partial^2 E_1\partial^2E_2\right],
\end{align} 
where, for simplicity, we have assumed $M_1$ and $M_2$ to be constants. We can see that different derivative interactions appear, including time and space derivatives. All these terms arise because of the non-zero value of $\alpha_{9}$ and $\alpha_{3,2}$ solely. 

It is important to mention that we have just shown one of the simplest cases that we can have with derivative interactions. The most general model satisfying the set of constraints given in eq.~(\ref{DerivativeCoeffsExample}) has a large number of free parameters, namely 19. This shows that there is a broad class of models with derivative interactions propagating only one scalar DoF at the linear level around homogeneous and isotropic backgrounds. Further restrictions on the parameters could be found by analysing the stability of the evolution of perturbations, as well as by looking for healthy non-linear completions. 

\bibliographystyle{apsrev4-1}
\bibliography{RefModifiedGravity}

\begin{thebibliography}{102}%
\makeatletter
\providecommand \@ifxundefined [1]{%
 \@ifx{#1\undefined}
}%
\providecommand \@ifnum [1]{%
 \ifnum #1\expandafter \@firstoftwo
 \else \expandafter \@secondoftwo
 \fi
}%
\providecommand \@ifx [1]{%
 \ifx #1\expandafter \@firstoftwo
 \else \expandafter \@secondoftwo
 \fi
}%
\providecommand \natexlab [1]{#1}%
\providecommand \enquote  [1]{``#1''}%
\providecommand \bibnamefont  [1]{#1}%
\providecommand \bibfnamefont [1]{#1}%
\providecommand \citenamefont [1]{#1}%
\providecommand \href@noop [0]{\@secondoftwo}%
\providecommand \href [0]{\begingroup \@sanitize@url \@href}%
\providecommand \@href[1]{\@@startlink{#1}\@@href}%
\providecommand \@@href[1]{\endgroup#1\@@endlink}%
\providecommand \@sanitize@url [0]{\catcode `\\12\catcode `\$12\catcode
  `\&12\catcode `\#12\catcode `\^12\catcode `\_12\catcode `\%12\relax}%
\providecommand \@@startlink[1]{}%
\providecommand \@@endlink[0]{}%
\providecommand \url  [0]{\begingroup\@sanitize@url \@url }%
\providecommand \@url [1]{\endgroup\@href {#1}{\urlprefix }}%
\providecommand \urlprefix  [0]{URL }%
\providecommand \Eprint [0]{\href }%
\providecommand \doibase [0]{http://dx.doi.org/}%
\providecommand \selectlanguage [0]{\@gobble}%
\providecommand \bibinfo  [0]{\@secondoftwo}%
\providecommand \bibfield  [0]{\@secondoftwo}%
\providecommand \translation [1]{[#1]}%
\providecommand \BibitemOpen [0]{}%
\providecommand \bibitemStop [0]{}%
\providecommand \bibitemNoStop [0]{.\EOS\space}%
\providecommand \EOS [0]{\spacefactor3000\relax}%
\providecommand \BibitemShut  [1]{\csname bibitem#1\endcsname}%
\let\auto@bib@innerbib\@empty
\bibitem [{\citenamefont {Lagos}\ \emph {et~al.}(2016)\citenamefont {Lagos},
  \citenamefont {Baker}, \citenamefont {Ferreira},\ and\ \citenamefont
  {Noller}}]{Lagos:2016wyv}%
  \BibitemOpen
  \bibfield  {author} {\bibinfo {author} {\bibfnamefont {M.}~\bibnamefont
  {Lagos}}, \bibinfo {author} {\bibfnamefont {T.}~\bibnamefont {Baker}},
  \bibinfo {author} {\bibfnamefont {P.~G.}\ \bibnamefont {Ferreira}}, \ and\
  \bibinfo {author} {\bibfnamefont {J.}~\bibnamefont {Noller}},\ }\href
  {\doibase 10.1088/1475-7516/2016/08/007} {\bibfield  {journal} {\bibinfo
  {journal} {JCAP}\ }\textbf {\bibinfo {volume} {1608}},\ \bibinfo {pages}
  {007} (\bibinfo {year} {2016})},\ \Eprint {http://arxiv.org/abs/1604.01396}
  {arXiv:1604.01396 [gr-qc]} \BibitemShut {NoStop}%
\bibitem [{\citenamefont {Creminelli}\ \emph {et~al.}(2009)\citenamefont
  {Creminelli}, \citenamefont {D'Amico}, \citenamefont {Norena},\ and\
  \citenamefont {Vernizzi}}]{Creminelli:2008wc}%
  \BibitemOpen
  \bibfield  {author} {\bibinfo {author} {\bibfnamefont {P.}~\bibnamefont
  {Creminelli}}, \bibinfo {author} {\bibfnamefont {G.}~\bibnamefont {D'Amico}},
  \bibinfo {author} {\bibfnamefont {J.}~\bibnamefont {Norena}}, \ and\ \bibinfo
  {author} {\bibfnamefont {F.}~\bibnamefont {Vernizzi}},\ }\href {\doibase
  10.1088/1475-7516/2009/02/018} {\bibfield  {journal} {\bibinfo  {journal}
  {JCAP}\ }\textbf {\bibinfo {volume} {0902}},\ \bibinfo {pages} {018}
  (\bibinfo {year} {2009})},\ \Eprint {http://arxiv.org/abs/0811.0827}
  {arXiv:0811.0827 [astro-ph]} \BibitemShut {NoStop}%
\bibitem [{\citenamefont {Zuntz}\ \emph {et~al.}(2012)\citenamefont {Zuntz},
  \citenamefont {Baker}, \citenamefont {Ferreira},\ and\ \citenamefont
  {Skordis}}]{Zuntz:2011aq}%
  \BibitemOpen
  \bibfield  {author} {\bibinfo {author} {\bibfnamefont {J.}~\bibnamefont
  {Zuntz}}, \bibinfo {author} {\bibfnamefont {T.}~\bibnamefont {Baker}},
  \bibinfo {author} {\bibfnamefont {P.}~\bibnamefont {Ferreira}}, \ and\
  \bibinfo {author} {\bibfnamefont {C.}~\bibnamefont {Skordis}},\ }\href
  {\doibase 10.1088/1475-7516/2012/06/032} {\bibfield  {journal} {\bibinfo
  {journal} {JCAP}\ }\textbf {\bibinfo {volume} {1206}},\ \bibinfo {pages}
  {032} (\bibinfo {year} {2012})},\ \Eprint {http://arxiv.org/abs/1110.3830}
  {arXiv:1110.3830 [astro-ph.CO]} \BibitemShut {NoStop}%
\bibitem [{\citenamefont {Baker}\ \emph {et~al.}(2013)\citenamefont {Baker},
  \citenamefont {Ferreira},\ and\ \citenamefont {Skordis}}]{Baker:2012zs}%
  \BibitemOpen
  \bibfield  {author} {\bibinfo {author} {\bibfnamefont {T.}~\bibnamefont
  {Baker}}, \bibinfo {author} {\bibfnamefont {P.~G.}\ \bibnamefont {Ferreira}},
  \ and\ \bibinfo {author} {\bibfnamefont {C.}~\bibnamefont {Skordis}},\ }\href
  {\doibase 10.1103/PhysRevD.87.024015} {\bibfield  {journal} {\bibinfo
  {journal} {Phys. Rev.}\ }\textbf {\bibinfo {volume} {D87}},\ \bibinfo {pages}
  {024015} (\bibinfo {year} {2013})},\ \Eprint {http://arxiv.org/abs/1209.2117}
  {arXiv:1209.2117 [astro-ph.CO]} \BibitemShut {NoStop}%
\bibitem [{\citenamefont {Gubitosi}\ \emph {et~al.}(2013)\citenamefont
  {Gubitosi}, \citenamefont {Piazza},\ and\ \citenamefont
  {Vernizzi}}]{Gubitosi:2012hu}%
  \BibitemOpen
  \bibfield  {author} {\bibinfo {author} {\bibfnamefont {G.}~\bibnamefont
  {Gubitosi}}, \bibinfo {author} {\bibfnamefont {F.}~\bibnamefont {Piazza}}, \
  and\ \bibinfo {author} {\bibfnamefont {F.}~\bibnamefont {Vernizzi}},\ }\href
  {\doibase 10.1088/1475-7516/2013/02/032} {\bibfield  {journal} {\bibinfo
  {journal} {JCAP}\ }\textbf {\bibinfo {volume} {1302}},\ \bibinfo {pages}
  {032} (\bibinfo {year} {2013})},\ \bibinfo {note} {[JCAP1302,032(2013)]},\
  \Eprint {http://arxiv.org/abs/1210.0201} {arXiv:1210.0201 [hep-th]}
  \BibitemShut {NoStop}%
\bibitem [{\citenamefont {Bloomfield}\ \emph {et~al.}(2013)\citenamefont
  {Bloomfield}, \citenamefont {Flanagan}, \citenamefont {Park},\ and\
  \citenamefont {Watson}}]{Bloomfield:2012ff}%
  \BibitemOpen
  \bibfield  {author} {\bibinfo {author} {\bibfnamefont {J.~K.}\ \bibnamefont
  {Bloomfield}}, \bibinfo {author} {\bibfnamefont {E.~E.}\ \bibnamefont
  {Flanagan}}, \bibinfo {author} {\bibfnamefont {M.}~\bibnamefont {Park}}, \
  and\ \bibinfo {author} {\bibfnamefont {S.}~\bibnamefont {Watson}},\ }\href
  {\doibase 10.1088/1475-7516/2013/08/010} {\bibfield  {journal} {\bibinfo
  {journal} {JCAP}\ }\textbf {\bibinfo {volume} {1308}},\ \bibinfo {pages}
  {010} (\bibinfo {year} {2013})},\ \Eprint {http://arxiv.org/abs/1211.7054}
  {arXiv:1211.7054 [astro-ph.CO]} \BibitemShut {NoStop}%
\bibitem [{\citenamefont {Gleyzes}\ \emph {et~al.}(2013)\citenamefont
  {Gleyzes}, \citenamefont {Langlois}, \citenamefont {Piazza},\ and\
  \citenamefont {Vernizzi}}]{Gleyzes:2013ooa}%
  \BibitemOpen
  \bibfield  {author} {\bibinfo {author} {\bibfnamefont {J.}~\bibnamefont
  {Gleyzes}}, \bibinfo {author} {\bibfnamefont {D.}~\bibnamefont {Langlois}},
  \bibinfo {author} {\bibfnamefont {F.}~\bibnamefont {Piazza}}, \ and\ \bibinfo
  {author} {\bibfnamefont {F.}~\bibnamefont {Vernizzi}},\ }\href {\doibase
  10.1088/1475-7516/2013/08/025} {\bibfield  {journal} {\bibinfo  {journal}
  {JCAP}\ }\textbf {\bibinfo {volume} {1308}},\ \bibinfo {pages} {025}
  (\bibinfo {year} {2013})},\ \Eprint {http://arxiv.org/abs/1304.4840}
  {arXiv:1304.4840 [hep-th]} \BibitemShut {NoStop}%
\bibitem [{\citenamefont {Bloomfield}(2013)}]{Bloomfield:2013efa}%
  \BibitemOpen
  \bibfield  {author} {\bibinfo {author} {\bibfnamefont {J.}~\bibnamefont
  {Bloomfield}},\ }\href {\doibase 10.1088/1475-7516/2013/12/044} {\bibfield
  {journal} {\bibinfo  {journal} {JCAP}\ }\textbf {\bibinfo {volume} {1312}},\
  \bibinfo {pages} {044} (\bibinfo {year} {2013})},\ \Eprint
  {http://arxiv.org/abs/1304.6712} {arXiv:1304.6712 [astro-ph.CO]} \BibitemShut
  {NoStop}%
\bibitem [{\citenamefont {Gleyzes}\ \emph {et~al.}(2015)\citenamefont
  {Gleyzes}, \citenamefont {Langlois},\ and\ \citenamefont
  {Vernizzi}}]{Gleyzes:2014rba}%
  \BibitemOpen
  \bibfield  {author} {\bibinfo {author} {\bibfnamefont {J.}~\bibnamefont
  {Gleyzes}}, \bibinfo {author} {\bibfnamefont {D.}~\bibnamefont {Langlois}}, \
  and\ \bibinfo {author} {\bibfnamefont {F.}~\bibnamefont {Vernizzi}},\ }\href
  {\doibase 10.1142/S021827181443010X} {\bibfield  {journal} {\bibinfo
  {journal} {Int. J. Mod. Phys.}\ }\textbf {\bibinfo {volume} {D23}},\ \bibinfo
  {pages} {1443010} (\bibinfo {year} {2015})},\ \Eprint
  {http://arxiv.org/abs/1411.3712} {arXiv:1411.3712 [hep-th]} \BibitemShut
  {NoStop}%
\bibitem [{\citenamefont {Battye}\ and\ \citenamefont
  {Pearson}(2012)}]{Battye:2012eu}%
  \BibitemOpen
  \bibfield  {author} {\bibinfo {author} {\bibfnamefont {R.~A.}\ \bibnamefont
  {Battye}}\ and\ \bibinfo {author} {\bibfnamefont {J.~A.}\ \bibnamefont
  {Pearson}},\ }\href {\doibase 10.1088/1475-7516/2012/07/019} {\bibfield
  {journal} {\bibinfo  {journal} {JCAP}\ }\textbf {\bibinfo {volume} {1207}},\
  \bibinfo {pages} {019} (\bibinfo {year} {2012})},\ \Eprint
  {http://arxiv.org/abs/1203.0398} {arXiv:1203.0398 [hep-th]} \BibitemShut
  {NoStop}%
\bibitem [{\citenamefont {Battye}\ and\ \citenamefont
  {Pearson}(2014)}]{Battye:2013ida}%
  \BibitemOpen
  \bibfield  {author} {\bibinfo {author} {\bibfnamefont {R.~A.}\ \bibnamefont
  {Battye}}\ and\ \bibinfo {author} {\bibfnamefont {J.~A.}\ \bibnamefont
  {Pearson}},\ }\href {\doibase 10.1088/1475-7516/2014/03/051} {\bibfield
  {journal} {\bibinfo  {journal} {JCAP}\ }\textbf {\bibinfo {volume} {1403}},\
  \bibinfo {pages} {051} (\bibinfo {year} {2014})},\ \Eprint
  {http://arxiv.org/abs/1311.6737} {arXiv:1311.6737 [astro-ph.CO]} \BibitemShut
  {NoStop}%
\bibitem [{\citenamefont {Skordis}\ \emph {et~al.}(2015)\citenamefont
  {Skordis}, \citenamefont {Pourtsidou},\ and\ \citenamefont
  {Copeland}}]{Skordis:2015yra}%
  \BibitemOpen
  \bibfield  {author} {\bibinfo {author} {\bibfnamefont {C.}~\bibnamefont
  {Skordis}}, \bibinfo {author} {\bibfnamefont {A.}~\bibnamefont {Pourtsidou}},
  \ and\ \bibinfo {author} {\bibfnamefont {E.~J.}\ \bibnamefont {Copeland}},\
  }\href {\doibase 10.1103/PhysRevD.91.083537} {\bibfield  {journal} {\bibinfo
  {journal} {Phys. Rev.}\ }\textbf {\bibinfo {volume} {D91}},\ \bibinfo {pages}
  {083537} (\bibinfo {year} {2015})},\ \Eprint
  {http://arxiv.org/abs/1502.07297} {arXiv:1502.07297 [astro-ph.CO]}
  \BibitemShut {NoStop}%
\bibitem [{\citenamefont {Volkov}(2012)}]{Volkov:2011an}%
  \BibitemOpen
  \bibfield  {author} {\bibinfo {author} {\bibfnamefont {M.~S.}\ \bibnamefont
  {Volkov}},\ }\href {\doibase 10.1007/JHEP01(2012)035} {\bibfield  {journal}
  {\bibinfo  {journal} {JHEP}\ }\textbf {\bibinfo {volume} {1201}},\ \bibinfo
  {pages} {035} (\bibinfo {year} {2012})},\ \Eprint
  {http://arxiv.org/abs/1110.6153} {arXiv:1110.6153 [hep-th]} \BibitemShut
  {NoStop}%
\bibitem [{\citenamefont {von Strauss}\ \emph {et~al.}(2012)\citenamefont {von
  Strauss}, \citenamefont {Schmidt-May}, \citenamefont {Enander}, \citenamefont
  {Mortsell},\ and\ \citenamefont {Hassan}}]{vonStrauss:2011mq}%
  \BibitemOpen
  \bibfield  {author} {\bibinfo {author} {\bibfnamefont {M.}~\bibnamefont {von
  Strauss}}, \bibinfo {author} {\bibfnamefont {A.}~\bibnamefont {Schmidt-May}},
  \bibinfo {author} {\bibfnamefont {J.}~\bibnamefont {Enander}}, \bibinfo
  {author} {\bibfnamefont {E.}~\bibnamefont {Mortsell}}, \ and\ \bibinfo
  {author} {\bibfnamefont {S.}~\bibnamefont {Hassan}},\ }\href {\doibase
  10.1088/1475-7516/2012/03/042} {\bibfield  {journal} {\bibinfo  {journal}
  {JCAP}\ }\textbf {\bibinfo {volume} {1203}},\ \bibinfo {pages} {042}
  (\bibinfo {year} {2012})},\ \Eprint {http://arxiv.org/abs/1111.1655}
  {arXiv:1111.1655 [gr-qc]} \BibitemShut {NoStop}%
\bibitem [{\citenamefont {Comelli}\ \emph
  {et~al.}(2012{\natexlab{a}})\citenamefont {Comelli}, \citenamefont
  {Crisostomi}, \citenamefont {Nesti},\ and\ \citenamefont
  {Pilo}}]{Comelli:2011zm}%
  \BibitemOpen
  \bibfield  {author} {\bibinfo {author} {\bibfnamefont {D.}~\bibnamefont
  {Comelli}}, \bibinfo {author} {\bibfnamefont {M.}~\bibnamefont {Crisostomi}},
  \bibinfo {author} {\bibfnamefont {F.}~\bibnamefont {Nesti}}, \ and\ \bibinfo
  {author} {\bibfnamefont {L.}~\bibnamefont {Pilo}},\ }\href {\doibase
  10.1007/JHEP06(2012)020, 10.1007/JHEP03(2012)067} {\bibfield  {journal}
  {\bibinfo  {journal} {JHEP}\ }\textbf {\bibinfo {volume} {1203}},\ \bibinfo
  {pages} {067} (\bibinfo {year} {2012}{\natexlab{a}})},\ \Eprint
  {http://arxiv.org/abs/1111.1983} {arXiv:1111.1983 [hep-th]} \BibitemShut
  {NoStop}%
\bibitem [{\citenamefont {Capozziello}\ and\ \citenamefont
  {Martin-Moruno}(2013)}]{Capozziello:2012re}%
  \BibitemOpen
  \bibfield  {author} {\bibinfo {author} {\bibfnamefont {S.}~\bibnamefont
  {Capozziello}}\ and\ \bibinfo {author} {\bibfnamefont {P.}~\bibnamefont
  {Martin-Moruno}},\ }\href {\doibase 10.1016/j.physletb.2013.01.025}
  {\bibfield  {journal} {\bibinfo  {journal} {Phys. Lett.}\ }\textbf {\bibinfo
  {volume} {B719}},\ \bibinfo {pages} {14} (\bibinfo {year} {2013})},\ \Eprint
  {http://arxiv.org/abs/1211.0214} {arXiv:1211.0214 [gr-qc]} \BibitemShut
  {NoStop}%
\bibitem [{\citenamefont {Comelli}\ \emph
  {et~al.}(2012{\natexlab{b}})\citenamefont {Comelli}, \citenamefont
  {Crisostomi},\ and\ \citenamefont {Pilo}}]{Comelli:2012db}%
  \BibitemOpen
  \bibfield  {author} {\bibinfo {author} {\bibfnamefont {D.}~\bibnamefont
  {Comelli}}, \bibinfo {author} {\bibfnamefont {M.}~\bibnamefont {Crisostomi}},
  \ and\ \bibinfo {author} {\bibfnamefont {L.}~\bibnamefont {Pilo}},\ }\href
  {\doibase 10.1007/JHEP06(2012)085} {\bibfield  {journal} {\bibinfo  {journal}
  {JHEP}\ }\textbf {\bibinfo {volume} {1206}},\ \bibinfo {pages} {085}
  (\bibinfo {year} {2012}{\natexlab{b}})},\ \Eprint
  {http://arxiv.org/abs/1202.1986} {arXiv:1202.1986 [hep-th]} \BibitemShut
  {NoStop}%
\bibitem [{\citenamefont {Berg}\ \emph {et~al.}(2012)\citenamefont {Berg},
  \citenamefont {Buchberger}, \citenamefont {Enander}, \citenamefont
  {Mortsell},\ and\ \citenamefont {Sjors}}]{Berg:2012kn}%
  \BibitemOpen
  \bibfield  {author} {\bibinfo {author} {\bibfnamefont {M.}~\bibnamefont
  {Berg}}, \bibinfo {author} {\bibfnamefont {I.}~\bibnamefont {Buchberger}},
  \bibinfo {author} {\bibfnamefont {J.}~\bibnamefont {Enander}}, \bibinfo
  {author} {\bibfnamefont {E.}~\bibnamefont {Mortsell}}, \ and\ \bibinfo
  {author} {\bibfnamefont {S.}~\bibnamefont {Sjors}},\ }\href {\doibase
  10.1088/1475-7516/2012/12/021} {\bibfield  {journal} {\bibinfo  {journal}
  {JCAP}\ }\textbf {\bibinfo {volume} {1212}},\ \bibinfo {pages} {021}
  (\bibinfo {year} {2012})},\ \Eprint {http://arxiv.org/abs/1206.3496}
  {arXiv:1206.3496 [gr-qc]} \BibitemShut {NoStop}%
\bibitem [{\citenamefont {Akrami}\ \emph
  {et~al.}(2013{\natexlab{a}})\citenamefont {Akrami}, \citenamefont
  {Koivisto},\ and\ \citenamefont {Sandstad}}]{Akrami:2012vf}%
  \BibitemOpen
  \bibfield  {author} {\bibinfo {author} {\bibfnamefont {Y.}~\bibnamefont
  {Akrami}}, \bibinfo {author} {\bibfnamefont {T.~S.}\ \bibnamefont
  {Koivisto}}, \ and\ \bibinfo {author} {\bibfnamefont {M.}~\bibnamefont
  {Sandstad}},\ }\href {\doibase 10.1007/JHEP03(2013)099} {\bibfield  {journal}
  {\bibinfo  {journal} {JHEP}\ }\textbf {\bibinfo {volume} {1303}},\ \bibinfo
  {pages} {099} (\bibinfo {year} {2013}{\natexlab{a}})},\ \Eprint
  {http://arxiv.org/abs/1209.0457} {arXiv:1209.0457 [astro-ph.CO]} \BibitemShut
  {NoStop}%
\bibitem [{\citenamefont {Sakakihara}\ \emph {et~al.}(2013)\citenamefont
  {Sakakihara}, \citenamefont {Soda},\ and\ \citenamefont
  {Takahashi}}]{Sakakihara:2012iq}%
  \BibitemOpen
  \bibfield  {author} {\bibinfo {author} {\bibfnamefont {Y.}~\bibnamefont
  {Sakakihara}}, \bibinfo {author} {\bibfnamefont {J.}~\bibnamefont {Soda}}, \
  and\ \bibinfo {author} {\bibfnamefont {T.}~\bibnamefont {Takahashi}},\ }\href
  {\doibase 10.1093/ptep/ptt004} {\bibfield  {journal} {\bibinfo  {journal}
  {PTEP}\ }\textbf {\bibinfo {volume} {2013}},\ \bibinfo {pages} {033E02}
  (\bibinfo {year} {2013})},\ \Eprint {http://arxiv.org/abs/1211.5976}
  {arXiv:1211.5976 [hep-th]} \BibitemShut {NoStop}%
\bibitem [{\citenamefont {Akrami}\ \emph
  {et~al.}(2013{\natexlab{b}})\citenamefont {Akrami}, \citenamefont {Koivisto},
  \citenamefont {Mota},\ and\ \citenamefont {Sandstad}}]{Akrami:2013ffa}%
  \BibitemOpen
  \bibfield  {author} {\bibinfo {author} {\bibfnamefont {Y.}~\bibnamefont
  {Akrami}}, \bibinfo {author} {\bibfnamefont {T.~S.}\ \bibnamefont
  {Koivisto}}, \bibinfo {author} {\bibfnamefont {D.~F.}\ \bibnamefont {Mota}},
  \ and\ \bibinfo {author} {\bibfnamefont {M.}~\bibnamefont {Sandstad}},\
  }\href {\doibase 10.1088/1475-7516/2013/10/046} {\bibfield  {journal}
  {\bibinfo  {journal} {JCAP}\ }\textbf {\bibinfo {volume} {1310}},\ \bibinfo
  {pages} {046} (\bibinfo {year} {2013}{\natexlab{b}})},\ \Eprint
  {http://arxiv.org/abs/1306.0004} {arXiv:1306.0004 [hep-th]} \BibitemShut
  {NoStop}%
\bibitem [{\citenamefont {Tamanini}\ \emph {et~al.}(2014)\citenamefont
  {Tamanini}, \citenamefont {Saridakis},\ and\ \citenamefont
  {Koivisto}}]{Tamanini:2013xia}%
  \BibitemOpen
  \bibfield  {author} {\bibinfo {author} {\bibfnamefont {N.}~\bibnamefont
  {Tamanini}}, \bibinfo {author} {\bibfnamefont {E.~N.}\ \bibnamefont
  {Saridakis}}, \ and\ \bibinfo {author} {\bibfnamefont {T.~S.}\ \bibnamefont
  {Koivisto}},\ }\href {\doibase 10.1088/1475-7516/2014/02/015} {\bibfield
  {journal} {\bibinfo  {journal} {JCAP}\ }\textbf {\bibinfo {volume} {1402}},\
  \bibinfo {pages} {015} (\bibinfo {year} {2014})},\ \Eprint
  {http://arxiv.org/abs/1307.5984} {arXiv:1307.5984 [hep-th]} \BibitemShut
  {NoStop}%
\bibitem [{\citenamefont {Aoki}\ and\ \citenamefont
  {Maeda}(2014{\natexlab{a}})}]{Aoki:2013joa}%
  \BibitemOpen
  \bibfield  {author} {\bibinfo {author} {\bibfnamefont {K.}~\bibnamefont
  {Aoki}}\ and\ \bibinfo {author} {\bibfnamefont {K.-i.}\ \bibnamefont
  {Maeda}},\ }\href {\doibase 10.1103/PhysRevD.89.064051} {\bibfield  {journal}
  {\bibinfo  {journal} {Phys.Rev.}\ }\textbf {\bibinfo {volume} {D89}},\
  \bibinfo {pages} {064051} (\bibinfo {year} {2014}{\natexlab{a}})},\ \Eprint
  {http://arxiv.org/abs/1312.7040} {arXiv:1312.7040 [gr-qc]} \BibitemShut
  {NoStop}%
\bibitem [{\citenamefont {Akrami}\ \emph
  {et~al.}(2015{\natexlab{a}})\citenamefont {Akrami}, \citenamefont
  {Koivisto},\ and\ \citenamefont {Sandstad}}]{Akrami:2013pna}%
  \BibitemOpen
  \bibfield  {author} {\bibinfo {author} {\bibfnamefont {Y.}~\bibnamefont
  {Akrami}}, \bibinfo {author} {\bibfnamefont {T.~S.}\ \bibnamefont
  {Koivisto}}, \ and\ \bibinfo {author} {\bibfnamefont {M.}~\bibnamefont
  {Sandstad}},\ }in\ \href {\doibase 10.1142/9789814623995_0138} {\emph
  {\bibinfo {booktitle} {{Proceedings, 13th Marcel Grossmann Meeting on Recent
  Developments in Theoretical and Experimental General Relativity,
  Astrophysics, and Relativistic Field Theories (MG13): Stockholm, Sweden, July
  1-7, 2012}}}}\ (\bibinfo {year} {2015})\ pp.\ \bibinfo {pages} {1252--1254},\
  \Eprint {http://arxiv.org/abs/1302.5268} {arXiv:1302.5268 [astro-ph.CO]}
  \BibitemShut {NoStop}%
\bibitem [{\citenamefont {De~Felice}\ \emph
  {et~al.}(2014{\natexlab{a}})\citenamefont {De~Felice}, \citenamefont
  {Nakamura},\ and\ \citenamefont {Tanaka}}]{DeFelice:2013nba}%
  \BibitemOpen
  \bibfield  {author} {\bibinfo {author} {\bibfnamefont {A.}~\bibnamefont
  {De~Felice}}, \bibinfo {author} {\bibfnamefont {T.}~\bibnamefont {Nakamura}},
  \ and\ \bibinfo {author} {\bibfnamefont {T.}~\bibnamefont {Tanaka}},\ }\href
  {\doibase 10.1093/ptep/ptu024} {\bibfield  {journal} {\bibinfo  {journal}
  {PTEP}\ }\textbf {\bibinfo {volume} {2014}},\ \bibinfo {pages} {043E01}
  (\bibinfo {year} {2014}{\natexlab{a}})},\ \Eprint
  {http://arxiv.org/abs/1304.3920} {arXiv:1304.3920 [gr-qc]} \BibitemShut
  {NoStop}%
\bibitem [{\citenamefont {Fasiello}\ and\ \citenamefont
  {Tolley}(2013)}]{Fasiello:2013woa}%
  \BibitemOpen
  \bibfield  {author} {\bibinfo {author} {\bibfnamefont {M.}~\bibnamefont
  {Fasiello}}\ and\ \bibinfo {author} {\bibfnamefont {A.~J.}\ \bibnamefont
  {Tolley}},\ }\href {\doibase 10.1088/1475-7516/2013/12/002} {\bibfield
  {journal} {\bibinfo  {journal} {JCAP}\ }\textbf {\bibinfo {volume} {1312}},\
  \bibinfo {pages} {002} (\bibinfo {year} {2013})},\ \Eprint
  {http://arxiv.org/abs/1308.1647} {arXiv:1308.1647 [hep-th]} \BibitemShut
  {NoStop}%
\bibitem [{\citenamefont {Volkov}(2013)}]{Volkov:2013roa}%
  \BibitemOpen
  \bibfield  {author} {\bibinfo {author} {\bibfnamefont {M.~S.}\ \bibnamefont
  {Volkov}},\ }\href {\doibase 10.1088/0264-9381/30/18/184009} {\bibfield
  {journal} {\bibinfo  {journal} {Class.Quant.Grav.}\ }\textbf {\bibinfo
  {volume} {30}},\ \bibinfo {pages} {184009} (\bibinfo {year} {2013})},\
  \Eprint {http://arxiv.org/abs/1304.0238} {arXiv:1304.0238 [hep-th]}
  \BibitemShut {NoStop}%
\bibitem [{\citenamefont {Koennig}\ \emph
  {et~al.}(2014{\natexlab{a}})\citenamefont {Koennig}, \citenamefont {Patil},\
  and\ \citenamefont {Amendola}}]{Konnig:2013gxa}%
  \BibitemOpen
  \bibfield  {author} {\bibinfo {author} {\bibfnamefont {F.}~\bibnamefont
  {Koennig}}, \bibinfo {author} {\bibfnamefont {A.}~\bibnamefont {Patil}}, \
  and\ \bibinfo {author} {\bibfnamefont {L.}~\bibnamefont {Amendola}},\ }\href
  {\doibase 10.1088/1475-7516/2014/03/029} {\bibfield  {journal} {\bibinfo
  {journal} {JCAP}\ }\textbf {\bibinfo {volume} {1403}},\ \bibinfo {pages}
  {029} (\bibinfo {year} {2014}{\natexlab{a}})},\ \Eprint
  {http://arxiv.org/abs/1312.3208} {arXiv:1312.3208 [astro-ph.CO]} \BibitemShut
  {NoStop}%
\bibitem [{\citenamefont {Berezhiani}\ \emph {et~al.}(2013)\citenamefont
  {Berezhiani}, \citenamefont {Chkareuli},\ and\ \citenamefont
  {Gabadadze}}]{Berezhiani:2013dw}%
  \BibitemOpen
  \bibfield  {author} {\bibinfo {author} {\bibfnamefont {L.}~\bibnamefont
  {Berezhiani}}, \bibinfo {author} {\bibfnamefont {G.}~\bibnamefont
  {Chkareuli}}, \ and\ \bibinfo {author} {\bibfnamefont {G.}~\bibnamefont
  {Gabadadze}},\ }\href {\doibase 10.1103/PhysRevD.88.124020} {\bibfield
  {journal} {\bibinfo  {journal} {Phys. Rev.}\ }\textbf {\bibinfo {volume}
  {D88}},\ \bibinfo {pages} {124020} (\bibinfo {year} {2013})},\ \Eprint
  {http://arxiv.org/abs/1302.0549} {arXiv:1302.0549 [hep-th]} \BibitemShut
  {NoStop}%
\bibitem [{\citenamefont {De~Felice}\ \emph
  {et~al.}(2014{\natexlab{b}})\citenamefont {De~Felice}, \citenamefont
  {Gumrukcuoglu}, \citenamefont {Mukohyama}, \citenamefont {Tanahashi},\ and\
  \citenamefont {Tanaka}}]{DeFelice:2014nja}%
  \BibitemOpen
  \bibfield  {author} {\bibinfo {author} {\bibfnamefont {A.}~\bibnamefont
  {De~Felice}}, \bibinfo {author} {\bibfnamefont {A.~E.}\ \bibnamefont
  {Gumrukcuoglu}}, \bibinfo {author} {\bibfnamefont {S.}~\bibnamefont
  {Mukohyama}}, \bibinfo {author} {\bibfnamefont {N.}~\bibnamefont
  {Tanahashi}}, \ and\ \bibinfo {author} {\bibfnamefont {T.}~\bibnamefont
  {Tanaka}},\ }\href {\doibase 10.1088/1475-7516/2014/06/037} {\bibfield
  {journal} {\bibinfo  {journal} {JCAP}\ }\textbf {\bibinfo {volume} {1406}},\
  \bibinfo {pages} {037} (\bibinfo {year} {2014}{\natexlab{b}})},\ \Eprint
  {http://arxiv.org/abs/1404.0008} {arXiv:1404.0008 [hep-th]} \BibitemShut
  {NoStop}%
\bibitem [{\citenamefont {de~Rham}\ \emph {et~al.}(2015)\citenamefont
  {de~Rham}, \citenamefont {Heisenberg},\ and\ \citenamefont
  {Ribeiro}}]{deRham:2014naa}%
  \BibitemOpen
  \bibfield  {author} {\bibinfo {author} {\bibfnamefont {C.}~\bibnamefont
  {de~Rham}}, \bibinfo {author} {\bibfnamefont {L.}~\bibnamefont {Heisenberg}},
  \ and\ \bibinfo {author} {\bibfnamefont {R.~H.}\ \bibnamefont {Ribeiro}},\
  }\href {\doibase 10.1088/0264-9381/32/3/035022} {\bibfield  {journal}
  {\bibinfo  {journal} {Class.Quant.Grav.}\ }\textbf {\bibinfo {volume} {32}},\
  \bibinfo {pages} {035022} (\bibinfo {year} {2015})},\ \Eprint
  {http://arxiv.org/abs/1408.1678} {arXiv:1408.1678 [hep-th]} \BibitemShut
  {NoStop}%
\bibitem [{\citenamefont {Enander}\ \emph
  {et~al.}(2015{\natexlab{a}})\citenamefont {Enander}, \citenamefont {Solomon},
  \citenamefont {Akrami},\ and\ \citenamefont {Mortsell}}]{Enander:2014xga}%
  \BibitemOpen
  \bibfield  {author} {\bibinfo {author} {\bibfnamefont {J.}~\bibnamefont
  {Enander}}, \bibinfo {author} {\bibfnamefont {A.~R.}\ \bibnamefont
  {Solomon}}, \bibinfo {author} {\bibfnamefont {Y.}~\bibnamefont {Akrami}}, \
  and\ \bibinfo {author} {\bibfnamefont {E.}~\bibnamefont {Mortsell}},\ }\href
  {\doibase 10.1088/1475-7516/2015/01/006} {\bibfield  {journal} {\bibinfo
  {journal} {JCAP}\ }\textbf {\bibinfo {volume} {1501}},\ \bibinfo {pages}
  {006} (\bibinfo {year} {2015}{\natexlab{a}})},\ \Eprint
  {http://arxiv.org/abs/1409.2860} {arXiv:1409.2860 [astro-ph.CO]} \BibitemShut
  {NoStop}%
\bibitem [{\citenamefont {Koennig}\ and\ \citenamefont
  {Amendola}(2014)}]{Konnig:2014dna}%
  \BibitemOpen
  \bibfield  {author} {\bibinfo {author} {\bibfnamefont {F.}~\bibnamefont
  {Koennig}}\ and\ \bibinfo {author} {\bibfnamefont {L.}~\bibnamefont
  {Amendola}},\ }\href {\doibase 10.1103/PhysRevD.90.044030} {\bibfield
  {journal} {\bibinfo  {journal} {Phys.Rev.}\ }\textbf {\bibinfo {volume}
  {D90}},\ \bibinfo {pages} {044030} (\bibinfo {year} {2014})},\ \Eprint
  {http://arxiv.org/abs/1402.1988} {arXiv:1402.1988 [astro-ph.CO]} \BibitemShut
  {NoStop}%
\bibitem [{\citenamefont {Solomon}\ \emph {et~al.}(2014)\citenamefont
  {Solomon}, \citenamefont {Akrami},\ and\ \citenamefont
  {Koivisto}}]{Solomon:2014dua}%
  \BibitemOpen
  \bibfield  {author} {\bibinfo {author} {\bibfnamefont {A.~R.}\ \bibnamefont
  {Solomon}}, \bibinfo {author} {\bibfnamefont {Y.}~\bibnamefont {Akrami}}, \
  and\ \bibinfo {author} {\bibfnamefont {T.~S.}\ \bibnamefont {Koivisto}},\
  }\href {\doibase 10.1088/1475-7516/2014/10/066} {\bibfield  {journal}
  {\bibinfo  {journal} {JCAP}\ }\textbf {\bibinfo {volume} {1410}},\ \bibinfo
  {pages} {066} (\bibinfo {year} {2014})},\ \Eprint
  {http://arxiv.org/abs/1404.4061} {arXiv:1404.4061 [astro-ph.CO]} \BibitemShut
  {NoStop}%
\bibitem [{\citenamefont {Comelli}\ \emph {et~al.}(2014)\citenamefont
  {Comelli}, \citenamefont {Crisostomi},\ and\ \citenamefont
  {Pilo}}]{Comelli:2014bqa}%
  \BibitemOpen
  \bibfield  {author} {\bibinfo {author} {\bibfnamefont {D.}~\bibnamefont
  {Comelli}}, \bibinfo {author} {\bibfnamefont {M.}~\bibnamefont {Crisostomi}},
  \ and\ \bibinfo {author} {\bibfnamefont {L.}~\bibnamefont {Pilo}},\ }\href
  {\doibase 10.1103/PhysRevD.90.084003} {\bibfield  {journal} {\bibinfo
  {journal} {Phys. Rev.}\ }\textbf {\bibinfo {volume} {D90}},\ \bibinfo {pages}
  {084003} (\bibinfo {year} {2014})},\ \Eprint {http://arxiv.org/abs/1403.5679}
  {arXiv:1403.5679 [hep-th]} \BibitemShut {NoStop}%
\bibitem [{\citenamefont {Koennig}\ \emph
  {et~al.}(2014{\natexlab{b}})\citenamefont {Koennig}, \citenamefont {Akrami},
  \citenamefont {Amendola}, \citenamefont {Motta},\ and\ \citenamefont
  {Solomon}}]{Konnig:2014xva}%
  \BibitemOpen
  \bibfield  {author} {\bibinfo {author} {\bibfnamefont {F.}~\bibnamefont
  {Koennig}}, \bibinfo {author} {\bibfnamefont {Y.}~\bibnamefont {Akrami}},
  \bibinfo {author} {\bibfnamefont {L.}~\bibnamefont {Amendola}}, \bibinfo
  {author} {\bibfnamefont {M.}~\bibnamefont {Motta}}, \ and\ \bibinfo {author}
  {\bibfnamefont {A.~R.}\ \bibnamefont {Solomon}},\ }\href {\doibase
  10.1103/PhysRevD.90.124014} {\bibfield  {journal} {\bibinfo  {journal} {Phys.
  Rev.}\ }\textbf {\bibinfo {volume} {D90}},\ \bibinfo {pages} {124014}
  (\bibinfo {year} {2014}{\natexlab{b}})},\ \Eprint
  {http://arxiv.org/abs/1407.4331} {arXiv:1407.4331 [astro-ph.CO]} \BibitemShut
  {NoStop}%
\bibitem [{\citenamefont {Lagos}\ and\ \citenamefont
  {Ferreira}(2014)}]{Lagos:2014lca}%
  \BibitemOpen
  \bibfield  {author} {\bibinfo {author} {\bibfnamefont {M.}~\bibnamefont
  {Lagos}}\ and\ \bibinfo {author} {\bibfnamefont {P.~G.}\ \bibnamefont
  {Ferreira}},\ }\href {\doibase 10.1088/1475-7516/2014/12/026} {\bibfield
  {journal} {\bibinfo  {journal} {JCAP}\ }\textbf {\bibinfo {volume} {1412}},\
  \bibinfo {pages} {026} (\bibinfo {year} {2014})},\ \Eprint
  {http://arxiv.org/abs/1410.0207} {arXiv:1410.0207 [gr-qc]} \BibitemShut
  {NoStop}%
\bibitem [{\citenamefont {Cusin}\ \emph
  {et~al.}(2015{\natexlab{a}})\citenamefont {Cusin}, \citenamefont {Durrer},
  \citenamefont {Guarato},\ and\ \citenamefont {Motta}}]{Cusin:2014psa}%
  \BibitemOpen
  \bibfield  {author} {\bibinfo {author} {\bibfnamefont {G.}~\bibnamefont
  {Cusin}}, \bibinfo {author} {\bibfnamefont {R.}~\bibnamefont {Durrer}},
  \bibinfo {author} {\bibfnamefont {P.}~\bibnamefont {Guarato}}, \ and\
  \bibinfo {author} {\bibfnamefont {M.}~\bibnamefont {Motta}},\ }\href
  {\doibase 10.1088/1475-7516/2015/05/030} {\bibfield  {journal} {\bibinfo
  {journal} {JCAP}\ }\textbf {\bibinfo {volume} {1505}},\ \bibinfo {pages}
  {030} (\bibinfo {year} {2015}{\natexlab{a}})},\ \Eprint
  {http://arxiv.org/abs/1412.5979} {arXiv:1412.5979 [astro-ph.CO]} \BibitemShut
  {NoStop}%
\bibitem [{\citenamefont {Emir~Gumrukcuoglu}\ \emph {et~al.}(2015)\citenamefont
  {Emir~Gumrukcuoglu}, \citenamefont {Heisenberg},\ and\ \citenamefont
  {Mukohyama}}]{Gumrukcuoglu:2014xba}%
  \BibitemOpen
  \bibfield  {author} {\bibinfo {author} {\bibfnamefont {A.}~\bibnamefont
  {Emir~Gumrukcuoglu}}, \bibinfo {author} {\bibfnamefont {L.}~\bibnamefont
  {Heisenberg}}, \ and\ \bibinfo {author} {\bibfnamefont {S.}~\bibnamefont
  {Mukohyama}},\ }\href {\doibase 10.1088/1475-7516/2015/02/022} {\bibfield
  {journal} {\bibinfo  {journal} {JCAP}\ }\textbf {\bibinfo {volume} {1502}},\
  \bibinfo {pages} {022} (\bibinfo {year} {2015})},\ \Eprint
  {http://arxiv.org/abs/1409.7260} {arXiv:1409.7260 [hep-th]} \BibitemShut
  {NoStop}%
\bibitem [{\citenamefont {Aoki}\ and\ \citenamefont
  {Maeda}(2014{\natexlab{b}})}]{Aoki:2014cla}%
  \BibitemOpen
  \bibfield  {author} {\bibinfo {author} {\bibfnamefont {K.}~\bibnamefont
  {Aoki}}\ and\ \bibinfo {author} {\bibfnamefont {K.-i.}\ \bibnamefont
  {Maeda}},\ }\href {\doibase 10.1103/PhysRevD.90.124089} {\bibfield  {journal}
  {\bibinfo  {journal} {Phys. Rev.}\ }\textbf {\bibinfo {volume} {D90}},\
  \bibinfo {pages} {124089} (\bibinfo {year} {2014}{\natexlab{b}})},\ \Eprint
  {http://arxiv.org/abs/1409.0202} {arXiv:1409.0202 [gr-qc]} \BibitemShut
  {NoStop}%
\bibitem [{\citenamefont {Enander}\ \emph
  {et~al.}(2015{\natexlab{b}})\citenamefont {Enander}, \citenamefont {Akrami},
  \citenamefont {Mortsell}, \citenamefont {Renneby},\ and\ \citenamefont
  {Solomon}}]{Enander:2015vja}%
  \BibitemOpen
  \bibfield  {author} {\bibinfo {author} {\bibfnamefont {J.}~\bibnamefont
  {Enander}}, \bibinfo {author} {\bibfnamefont {Y.}~\bibnamefont {Akrami}},
  \bibinfo {author} {\bibfnamefont {E.}~\bibnamefont {Mortsell}}, \bibinfo
  {author} {\bibfnamefont {M.}~\bibnamefont {Renneby}}, \ and\ \bibinfo
  {author} {\bibfnamefont {A.~R.}\ \bibnamefont {Solomon}},\ }\href {\doibase
  10.1103/PhysRevD.91.084046} {\bibfield  {journal} {\bibinfo  {journal}
  {Phys.Rev.}\ }\textbf {\bibinfo {volume} {D91}},\ \bibinfo {pages} {084046}
  (\bibinfo {year} {2015}{\natexlab{b}})},\ \Eprint
  {http://arxiv.org/abs/1501.02140} {arXiv:1501.02140 [astro-ph.CO]}
  \BibitemShut {NoStop}%
\bibitem [{\citenamefont {Nersisyan}\ \emph {et~al.}(2015)\citenamefont
  {Nersisyan}, \citenamefont {Akrami},\ and\ \citenamefont
  {Amendola}}]{Nersisyan:2015oha}%
  \BibitemOpen
  \bibfield  {author} {\bibinfo {author} {\bibfnamefont {H.}~\bibnamefont
  {Nersisyan}}, \bibinfo {author} {\bibfnamefont {Y.}~\bibnamefont {Akrami}}, \
  and\ \bibinfo {author} {\bibfnamefont {L.}~\bibnamefont {Amendola}},\ }\href
  {\doibase 10.1103/PhysRevD.92.104034} {\bibfield  {journal} {\bibinfo
  {journal} {Phys. Rev.}\ }\textbf {\bibinfo {volume} {D92}},\ \bibinfo {pages}
  {104034} (\bibinfo {year} {2015})},\ \Eprint
  {http://arxiv.org/abs/1502.03988} {arXiv:1502.03988 [gr-qc]} \BibitemShut
  {NoStop}%
\bibitem [{\citenamefont {Soloviev}(2015)}]{Soloviev:2015wya}%
  \BibitemOpen
  \bibfield  {author} {\bibinfo {author} {\bibfnamefont {V.~O.}\ \bibnamefont
  {Soloviev}},\ }\href@noop {} {\  (\bibinfo {year} {2015})},\ \Eprint
  {http://arxiv.org/abs/1505.00840} {arXiv:1505.00840 [hep-th]} \BibitemShut
  {NoStop}%
\bibitem [{\citenamefont {Amendola}\ \emph {et~al.}(2015)\citenamefont
  {Amendola}, \citenamefont {Koennig}, \citenamefont {Martinelli},
  \citenamefont {Pettorino},\ and\ \citenamefont
  {Zumalacarregui}}]{Amendola:2015tua}%
  \BibitemOpen
  \bibfield  {author} {\bibinfo {author} {\bibfnamefont {L.}~\bibnamefont
  {Amendola}}, \bibinfo {author} {\bibfnamefont {F.}~\bibnamefont {Koennig}},
  \bibinfo {author} {\bibfnamefont {M.}~\bibnamefont {Martinelli}}, \bibinfo
  {author} {\bibfnamefont {V.}~\bibnamefont {Pettorino}}, \ and\ \bibinfo
  {author} {\bibfnamefont {M.}~\bibnamefont {Zumalacarregui}},\ }\href
  {\doibase 10.1088/1475-7516/2015/05/052} {\bibfield  {journal} {\bibinfo
  {journal} {JCAP}\ }\textbf {\bibinfo {volume} {1505}},\ \bibinfo {pages}
  {052} (\bibinfo {year} {2015})},\ \Eprint {http://arxiv.org/abs/1503.02490}
  {arXiv:1503.02490 [astro-ph.CO]} \BibitemShut {NoStop}%
\bibitem [{\citenamefont {Johnson}\ and\ \citenamefont
  {Terrana}(2015)}]{Johnson:2015tfa}%
  \BibitemOpen
  \bibfield  {author} {\bibinfo {author} {\bibfnamefont {M.}~\bibnamefont
  {Johnson}}\ and\ \bibinfo {author} {\bibfnamefont {A.}~\bibnamefont
  {Terrana}},\ }\href {\doibase 10.1103/PhysRevD.92.044001} {\bibfield
  {journal} {\bibinfo  {journal} {Phys. Rev.}\ }\textbf {\bibinfo {volume}
  {D92}},\ \bibinfo {pages} {044001} (\bibinfo {year} {2015})},\ \Eprint
  {http://arxiv.org/abs/1503.05560} {arXiv:1503.05560 [astro-ph.CO]}
  \BibitemShut {NoStop}%
\bibitem [{\citenamefont {Koennig}(2015)}]{Konnig:2015lfa}%
  \BibitemOpen
  \bibfield  {author} {\bibinfo {author} {\bibfnamefont {F.}~\bibnamefont
  {Koennig}},\ }\href {\doibase 10.1103/PhysRevD.91.104019} {\bibfield
  {journal} {\bibinfo  {journal} {Phys. Rev.}\ }\textbf {\bibinfo {volume}
  {D91}},\ \bibinfo {pages} {104019} (\bibinfo {year} {2015})},\ \Eprint
  {http://arxiv.org/abs/1503.07436} {arXiv:1503.07436 [astro-ph.CO]}
  \BibitemShut {NoStop}%
\bibitem [{\citenamefont {Cusin}\ \emph
  {et~al.}(2015{\natexlab{b}})\citenamefont {Cusin}, \citenamefont {Durrer},
  \citenamefont {Guarato},\ and\ \citenamefont {Motta}}]{Cusin:2015pya}%
  \BibitemOpen
  \bibfield  {author} {\bibinfo {author} {\bibfnamefont {G.}~\bibnamefont
  {Cusin}}, \bibinfo {author} {\bibfnamefont {R.}~\bibnamefont {Durrer}},
  \bibinfo {author} {\bibfnamefont {P.}~\bibnamefont {Guarato}}, \ and\
  \bibinfo {author} {\bibfnamefont {M.}~\bibnamefont {Motta}},\ }\href
  {\doibase 10.1088/1475-7516/2015/09/043, 10.1088/1475-7516/2015/9/043}
  {\bibfield  {journal} {\bibinfo  {journal} {JCAP}\ }\textbf {\bibinfo
  {volume} {1509}},\ \bibinfo {pages} {043} (\bibinfo {year}
  {2015}{\natexlab{b}})},\ \Eprint {http://arxiv.org/abs/1505.01091}
  {arXiv:1505.01091 [astro-ph.CO]} \BibitemShut {NoStop}%
\bibitem [{\citenamefont {Fasiello}\ and\ \citenamefont
  {Ribeiro}(2015)}]{Fasiello:2015csa}%
  \BibitemOpen
  \bibfield  {author} {\bibinfo {author} {\bibfnamefont {M.}~\bibnamefont
  {Fasiello}}\ and\ \bibinfo {author} {\bibfnamefont {R.~H.}\ \bibnamefont
  {Ribeiro}},\ }\href {\doibase 10.1088/1475-7516/2015/07/027} {\bibfield
  {journal} {\bibinfo  {journal} {JCAP}\ }\textbf {\bibinfo {volume} {1507}},\
  \bibinfo {pages} {027} (\bibinfo {year} {2015})},\ \Eprint
  {http://arxiv.org/abs/1505.00404} {arXiv:1505.00404 [astro-ph.CO]}
  \BibitemShut {NoStop}%
\bibitem [{\citenamefont {Akrami}\ \emph
  {et~al.}(2015{\natexlab{b}})\citenamefont {Akrami}, \citenamefont {Hassan},
  \citenamefont {Koennig}, \citenamefont {Schmidt-May},\ and\ \citenamefont
  {Solomon}}]{Akrami:2015qga}%
  \BibitemOpen
  \bibfield  {author} {\bibinfo {author} {\bibfnamefont {Y.}~\bibnamefont
  {Akrami}}, \bibinfo {author} {\bibfnamefont {S.~F.}\ \bibnamefont {Hassan}},
  \bibinfo {author} {\bibfnamefont {F.}~\bibnamefont {Koennig}}, \bibinfo
  {author} {\bibfnamefont {A.}~\bibnamefont {Schmidt-May}}, \ and\ \bibinfo
  {author} {\bibfnamefont {A.~R.}\ \bibnamefont {Solomon}},\ }\href {\doibase
  10.1016/j.physletb.2015.06.062} {\bibfield  {journal} {\bibinfo  {journal}
  {Phys. Lett.}\ }\textbf {\bibinfo {volume} {B748}},\ \bibinfo {pages} {37}
  (\bibinfo {year} {2015}{\natexlab{b}})},\ \Eprint
  {http://arxiv.org/abs/1503.07521} {arXiv:1503.07521 [gr-qc]} \BibitemShut
  {NoStop}%
\bibitem [{\citenamefont {Aoki}\ \emph {et~al.}(2015)\citenamefont {Aoki},
  \citenamefont {Maeda},\ and\ \citenamefont {Namba}}]{Aoki:2015xqa}%
  \BibitemOpen
  \bibfield  {author} {\bibinfo {author} {\bibfnamefont {K.}~\bibnamefont
  {Aoki}}, \bibinfo {author} {\bibfnamefont {K.-i.}\ \bibnamefont {Maeda}}, \
  and\ \bibinfo {author} {\bibfnamefont {R.}~\bibnamefont {Namba}},\ }\href
  {\doibase 10.1103/PhysRevD.92.044054} {\bibfield  {journal} {\bibinfo
  {journal} {Phys. Rev.}\ }\textbf {\bibinfo {volume} {D92}},\ \bibinfo {pages}
  {044054} (\bibinfo {year} {2015})},\ \Eprint
  {http://arxiv.org/abs/1506.04543} {arXiv:1506.04543 [hep-th]} \BibitemShut
  {NoStop}%
\bibitem [{\citenamefont {Lagos}\ and\ \citenamefont
  {Noller}(2016)}]{Lagos:2015sya}%
  \BibitemOpen
  \bibfield  {author} {\bibinfo {author} {\bibfnamefont {M.}~\bibnamefont
  {Lagos}}\ and\ \bibinfo {author} {\bibfnamefont {J.}~\bibnamefont {Noller}},\
  }\href {\doibase 10.1088/1475-7516/2016/01/023} {\bibfield  {journal}
  {\bibinfo  {journal} {JCAP}\ }\textbf {\bibinfo {volume} {1601}},\ \bibinfo
  {pages} {023} (\bibinfo {year} {2016})},\ \Eprint
  {http://arxiv.org/abs/1508.05864} {arXiv:1508.05864 [gr-qc]} \BibitemShut
  {NoStop}%
\bibitem [{\citenamefont {Cusin}\ \emph {et~al.}(2016)\citenamefont {Cusin},
  \citenamefont {Durrer}, \citenamefont {Guarato},\ and\ \citenamefont
  {Motta}}]{Cusin:2015tmf}%
  \BibitemOpen
  \bibfield  {author} {\bibinfo {author} {\bibfnamefont {G.}~\bibnamefont
  {Cusin}}, \bibinfo {author} {\bibfnamefont {R.}~\bibnamefont {Durrer}},
  \bibinfo {author} {\bibfnamefont {P.}~\bibnamefont {Guarato}}, \ and\
  \bibinfo {author} {\bibfnamefont {M.}~\bibnamefont {Motta}},\ }\href
  {\doibase 10.1088/1475-7516/2016/04/051} {\bibfield  {journal} {\bibinfo
  {journal} {JCAP}\ }\textbf {\bibinfo {volume} {1604}},\ \bibinfo {pages}
  {051} (\bibinfo {year} {2016})},\ \Eprint {http://arxiv.org/abs/1512.02131}
  {arXiv:1512.02131 [astro-ph.CO]} \BibitemShut {NoStop}%
\bibitem [{\citenamefont {Comelli}\ \emph {et~al.}(2015)\citenamefont
  {Comelli}, \citenamefont {Crisostomi}, \citenamefont {Koyama}, \citenamefont
  {Pilo},\ and\ \citenamefont {Tasinato}}]{Comelli:2015pua}%
  \BibitemOpen
  \bibfield  {author} {\bibinfo {author} {\bibfnamefont {D.}~\bibnamefont
  {Comelli}}, \bibinfo {author} {\bibfnamefont {M.}~\bibnamefont {Crisostomi}},
  \bibinfo {author} {\bibfnamefont {K.}~\bibnamefont {Koyama}}, \bibinfo
  {author} {\bibfnamefont {L.}~\bibnamefont {Pilo}}, \ and\ \bibinfo {author}
  {\bibfnamefont {G.}~\bibnamefont {Tasinato}},\ }\href {\doibase
  10.1088/1475-7516/2015/04/026} {\bibfield  {journal} {\bibinfo  {journal}
  {JCAP}\ }\textbf {\bibinfo {volume} {1504}},\ \bibinfo {pages} {026}
  (\bibinfo {year} {2015})},\ \Eprint {http://arxiv.org/abs/1501.00864}
  {arXiv:1501.00864 [hep-th]} \BibitemShut {NoStop}%
\bibitem [{\citenamefont {Heisenberg}\ and\ \citenamefont
  {Refregier}(2016)}]{Heisenberg:2016spl}%
  \BibitemOpen
  \bibfield  {author} {\bibinfo {author} {\bibfnamefont {L.}~\bibnamefont
  {Heisenberg}}\ and\ \bibinfo {author} {\bibfnamefont {A.}~\bibnamefont
  {Refregier}},\ }\href {\doibase 10.1088/1475-7516/2016/09/020} {\bibfield
  {journal} {\bibinfo  {journal} {JCAP}\ }\textbf {\bibinfo {volume} {1609}},\
  \bibinfo {pages} {020} (\bibinfo {year} {2016})},\ \Eprint
  {http://arxiv.org/abs/1604.07306} {arXiv:1604.07306 [gr-qc]} \BibitemShut
  {NoStop}%
\bibitem [{\citenamefont {de~Rham}\ \emph
  {et~al.}(2011{\natexlab{a}})\citenamefont {de~Rham}, \citenamefont
  {Gabadadze}, \citenamefont {Heisenberg},\ and\ \citenamefont
  {Pirtskhalava}}]{deRham:2010tw}%
  \BibitemOpen
  \bibfield  {author} {\bibinfo {author} {\bibfnamefont {C.}~\bibnamefont
  {de~Rham}}, \bibinfo {author} {\bibfnamefont {G.}~\bibnamefont {Gabadadze}},
  \bibinfo {author} {\bibfnamefont {L.}~\bibnamefont {Heisenberg}}, \ and\
  \bibinfo {author} {\bibfnamefont {D.}~\bibnamefont {Pirtskhalava}},\ }\href
  {\doibase 10.1103/PhysRevD.83.103516} {\bibfield  {journal} {\bibinfo
  {journal} {Phys. Rev.}\ }\textbf {\bibinfo {volume} {D83}},\ \bibinfo {pages}
  {103516} (\bibinfo {year} {2011}{\natexlab{a}})},\ \Eprint
  {http://arxiv.org/abs/1010.1780} {arXiv:1010.1780 [hep-th]} \BibitemShut
  {NoStop}%
\bibitem [{\citenamefont {D'Amico}\ \emph {et~al.}(2011)\citenamefont
  {D'Amico}, \citenamefont {de~Rham}, \citenamefont {Dubovsky}, \citenamefont
  {Gabadadze}, \citenamefont {Pirtskhalava},\ and\ \citenamefont
  {Tolley}}]{D'Amico:2011jj}%
  \BibitemOpen
  \bibfield  {author} {\bibinfo {author} {\bibfnamefont {G.}~\bibnamefont
  {D'Amico}}, \bibinfo {author} {\bibfnamefont {C.}~\bibnamefont {de~Rham}},
  \bibinfo {author} {\bibfnamefont {S.}~\bibnamefont {Dubovsky}}, \bibinfo
  {author} {\bibfnamefont {G.}~\bibnamefont {Gabadadze}}, \bibinfo {author}
  {\bibfnamefont {D.}~\bibnamefont {Pirtskhalava}}, \ and\ \bibinfo {author}
  {\bibfnamefont {A.~J.}\ \bibnamefont {Tolley}},\ }\href {\doibase
  10.1103/PhysRevD.84.124046} {\bibfield  {journal} {\bibinfo  {journal} {Phys.
  Rev.}\ }\textbf {\bibinfo {volume} {D84}},\ \bibinfo {pages} {124046}
  (\bibinfo {year} {2011})},\ \Eprint {http://arxiv.org/abs/1108.5231}
  {arXiv:1108.5231 [hep-th]} \BibitemShut {NoStop}%
\bibitem [{\citenamefont {Gumrukcuoglu}\ \emph {et~al.}(2011)\citenamefont
  {Gumrukcuoglu}, \citenamefont {Lin},\ and\ \citenamefont
  {Mukohyama}}]{Gumrukcuoglu:2011ew}%
  \BibitemOpen
  \bibfield  {author} {\bibinfo {author} {\bibfnamefont {A.~E.}\ \bibnamefont
  {Gumrukcuoglu}}, \bibinfo {author} {\bibfnamefont {C.}~\bibnamefont {Lin}}, \
  and\ \bibinfo {author} {\bibfnamefont {S.}~\bibnamefont {Mukohyama}},\ }\href
  {\doibase 10.1088/1475-7516/2011/11/030} {\bibfield  {journal} {\bibinfo
  {journal} {JCAP}\ }\textbf {\bibinfo {volume} {1111}},\ \bibinfo {pages}
  {030} (\bibinfo {year} {2011})},\ \Eprint {http://arxiv.org/abs/1109.3845}
  {arXiv:1109.3845 [hep-th]} \BibitemShut {NoStop}%
\bibitem [{\citenamefont {Gumrukcuoglu}\ \emph {et~al.}(2012)\citenamefont
  {Gumrukcuoglu}, \citenamefont {Lin},\ and\ \citenamefont
  {Mukohyama}}]{Gumrukcuoglu:2011zh}%
  \BibitemOpen
  \bibfield  {author} {\bibinfo {author} {\bibfnamefont {A.~E.}\ \bibnamefont
  {Gumrukcuoglu}}, \bibinfo {author} {\bibfnamefont {C.}~\bibnamefont {Lin}}, \
  and\ \bibinfo {author} {\bibfnamefont {S.}~\bibnamefont {Mukohyama}},\ }\href
  {\doibase 10.1088/1475-7516/2012/03/006} {\bibfield  {journal} {\bibinfo
  {journal} {JCAP}\ }\textbf {\bibinfo {volume} {1203}},\ \bibinfo {pages}
  {006} (\bibinfo {year} {2012})},\ \Eprint {http://arxiv.org/abs/1111.4107}
  {arXiv:1111.4107 [hep-th]} \BibitemShut {NoStop}%
\bibitem [{\citenamefont {Vakili}\ and\ \citenamefont
  {Khosravi}(2012)}]{Vakili:2012tm}%
  \BibitemOpen
  \bibfield  {author} {\bibinfo {author} {\bibfnamefont {B.}~\bibnamefont
  {Vakili}}\ and\ \bibinfo {author} {\bibfnamefont {N.}~\bibnamefont
  {Khosravi}},\ }\href {\doibase 10.1103/PhysRevD.85.083529} {\bibfield
  {journal} {\bibinfo  {journal} {Phys. Rev.}\ }\textbf {\bibinfo {volume}
  {D85}},\ \bibinfo {pages} {083529} (\bibinfo {year} {2012})},\ \Eprint
  {http://arxiv.org/abs/1204.1456} {arXiv:1204.1456 [gr-qc]} \BibitemShut
  {NoStop}%
\bibitem [{\citenamefont {De~Felice}\ \emph {et~al.}(2012)\citenamefont
  {De~Felice}, \citenamefont {Gumrukcuoglu},\ and\ \citenamefont
  {Mukohyama}}]{DeFelice:2012mx}%
  \BibitemOpen
  \bibfield  {author} {\bibinfo {author} {\bibfnamefont {A.}~\bibnamefont
  {De~Felice}}, \bibinfo {author} {\bibfnamefont {A.~E.}\ \bibnamefont
  {Gumrukcuoglu}}, \ and\ \bibinfo {author} {\bibfnamefont {S.}~\bibnamefont
  {Mukohyama}},\ }\href {\doibase 10.1103/PhysRevLett.109.171101} {\bibfield
  {journal} {\bibinfo  {journal} {Phys. Rev. Lett.}\ }\textbf {\bibinfo
  {volume} {109}},\ \bibinfo {pages} {171101} (\bibinfo {year} {2012})},\
  \Eprint {http://arxiv.org/abs/1206.2080} {arXiv:1206.2080 [hep-th]}
  \BibitemShut {NoStop}%
\bibitem [{\citenamefont {Fasiello}\ and\ \citenamefont
  {Tolley}(2012)}]{Fasiello:2012rw}%
  \BibitemOpen
  \bibfield  {author} {\bibinfo {author} {\bibfnamefont {M.}~\bibnamefont
  {Fasiello}}\ and\ \bibinfo {author} {\bibfnamefont {A.~J.}\ \bibnamefont
  {Tolley}},\ }\href {\doibase 10.1088/1475-7516/2012/11/035} {\bibfield
  {journal} {\bibinfo  {journal} {JCAP}\ }\textbf {\bibinfo {volume} {1211}},\
  \bibinfo {pages} {035} (\bibinfo {year} {2012})},\ \Eprint
  {http://arxiv.org/abs/1206.3852} {arXiv:1206.3852 [hep-th]} \BibitemShut
  {NoStop}%
\bibitem [{\citenamefont {de~Rham}\ \emph
  {et~al.}(2014{\natexlab{a}})\citenamefont {de~Rham}, \citenamefont
  {Fasiello},\ and\ \citenamefont {Tolley}}]{deRham:2014gla}%
  \BibitemOpen
  \bibfield  {author} {\bibinfo {author} {\bibfnamefont {C.}~\bibnamefont
  {de~Rham}}, \bibinfo {author} {\bibfnamefont {M.}~\bibnamefont {Fasiello}}, \
  and\ \bibinfo {author} {\bibfnamefont {A.~J.}\ \bibnamefont {Tolley}},\
  }\href {\doibase 10.1142/S0218271814430068} {\bibfield  {journal} {\bibinfo
  {journal} {Int. J. Mod. Phys.}\ }\textbf {\bibinfo {volume} {D23}},\ \bibinfo
  {pages} {1443006} (\bibinfo {year} {2014}{\natexlab{a}})},\ \Eprint
  {http://arxiv.org/abs/1410.0960} {arXiv:1410.0960 [hep-th]} \BibitemShut
  {NoStop}%
\bibitem [{\citenamefont {Pereira}\ \emph {et~al.}(2016)\citenamefont
  {Pereira}, \citenamefont {Mendonça}, \citenamefont {S.},\ and\ \citenamefont
  {Jesus}}]{Pereira:2015jua}%
  \BibitemOpen
  \bibfield  {author} {\bibinfo {author} {\bibfnamefont {S.~H.}\ \bibnamefont
  {Pereira}}, \bibinfo {author} {\bibfnamefont {E.~L.}\ \bibnamefont
  {Mendonça}}, \bibinfo {author} {\bibfnamefont {A.~P.~S.}\ \bibnamefont
  {S.}}, \ and\ \bibinfo {author} {\bibfnamefont {J.~F.}\ \bibnamefont
  {Jesus}},\ }\href@noop {} {\bibfield  {journal} {\bibinfo  {journal} {Rev.
  Mex. Astron. Astrofis.}\ }\textbf {\bibinfo {volume} {52}},\ \bibinfo {pages}
  {125} (\bibinfo {year} {2016})},\ \Eprint {http://arxiv.org/abs/1504.02295}
  {arXiv:1504.02295 [gr-qc]} \BibitemShut {NoStop}%
\bibitem [{\citenamefont {Banados}\ and\ \citenamefont
  {Ferreira}(2010)}]{Banados:2010ix}%
  \BibitemOpen
  \bibfield  {author} {\bibinfo {author} {\bibfnamefont {M.}~\bibnamefont
  {Banados}}\ and\ \bibinfo {author} {\bibfnamefont {P.~G.}\ \bibnamefont
  {Ferreira}},\ }\href {\doibase 10.1103/PhysRevLett.105.011101} {\bibfield
  {journal} {\bibinfo  {journal} {Phys.Rev.Lett.}\ }\textbf {\bibinfo {volume}
  {105}},\ \bibinfo {pages} {011101} (\bibinfo {year} {2010})},\ \Eprint
  {http://arxiv.org/abs/1006.1769} {arXiv:1006.1769 [astro-ph.CO]} \BibitemShut
  {NoStop}%
\bibitem [{\citenamefont {Scargill}\ \emph {et~al.}(2012)\citenamefont
  {Scargill}, \citenamefont {Banados},\ and\ \citenamefont
  {Ferreira}}]{Scargill:2012kg}%
  \BibitemOpen
  \bibfield  {author} {\bibinfo {author} {\bibfnamefont {J.~H.~C.}\
  \bibnamefont {Scargill}}, \bibinfo {author} {\bibfnamefont {M.}~\bibnamefont
  {Banados}}, \ and\ \bibinfo {author} {\bibfnamefont {P.~G.}\ \bibnamefont
  {Ferreira}},\ }\href {\doibase 10.1103/PhysRevD.86.103533} {\bibfield
  {journal} {\bibinfo  {journal} {Phys. Rev.}\ }\textbf {\bibinfo {volume}
  {D86}},\ \bibinfo {pages} {103533} (\bibinfo {year} {2012})},\ \Eprint
  {http://arxiv.org/abs/1210.1521} {arXiv:1210.1521 [astro-ph.CO]} \BibitemShut
  {NoStop}%
\bibitem [{\citenamefont {Escamilla-Rivera}\ \emph {et~al.}(2015)\citenamefont
  {Escamilla-Rivera}, \citenamefont {Banados},\ and\ \citenamefont
  {Ferreira}}]{EscamillaRivera:2013hv}%
  \BibitemOpen
  \bibfield  {author} {\bibinfo {author} {\bibfnamefont {C.}~\bibnamefont
  {Escamilla-Rivera}}, \bibinfo {author} {\bibfnamefont {M.}~\bibnamefont
  {Banados}}, \ and\ \bibinfo {author} {\bibfnamefont {P.~G.}\ \bibnamefont
  {Ferreira}},\ }in\ \href {\doibase 10.1142/9789814623995_0157} {\emph
  {\bibinfo {booktitle} {{Proceedings, 13th Marcel Grossmann Meeting on Recent
  Developments in Theoretical and Experimental General Relativity,
  Astrophysics, and Relativistic Field Theories (MG13): Stockholm, Sweden, July
  1-7, 2012}}}}\ (\bibinfo {year} {2015})\ pp.\ \bibinfo {pages} {1310--1312},\
  \Eprint {http://arxiv.org/abs/1301.5264} {arXiv:1301.5264 [gr-qc]}
  \BibitemShut {NoStop}%
\bibitem [{\citenamefont {Lagos}\ \emph {et~al.}(2014)\citenamefont {Lagos},
  \citenamefont {Ba\~nados}, \citenamefont {Ferreira},\ and\ \citenamefont
  {Garc\'ia-S\'aenz}}]{PhysRevD.89.024034}%
  \BibitemOpen
  \bibfield  {author} {\bibinfo {author} {\bibfnamefont {M.}~\bibnamefont
  {Lagos}}, \bibinfo {author} {\bibfnamefont {M.}~\bibnamefont {Ba\~nados}},
  \bibinfo {author} {\bibfnamefont {P.~G.}\ \bibnamefont {Ferreira}}, \ and\
  \bibinfo {author} {\bibfnamefont {S.}~\bibnamefont {Garc\'ia-S\'aenz}},\
  }\href {\doibase 10.1103/PhysRevD.89.024034} {\bibfield  {journal} {\bibinfo
  {journal} {Phys. Rev. D}\ }\textbf {\bibinfo {volume} {89}},\ \bibinfo
  {pages} {024034} (\bibinfo {year} {2014})}\BibitemShut {NoStop}%
\bibitem [{\citenamefont {Bouhmadi-Lopez}\ \emph {et~al.}(2014)\citenamefont
  {Bouhmadi-Lopez}, \citenamefont {Chen},\ and\ \citenamefont
  {Chen}}]{Bouhmadi-Lopez:2013lha}%
  \BibitemOpen
  \bibfield  {author} {\bibinfo {author} {\bibfnamefont {M.}~\bibnamefont
  {Bouhmadi-Lopez}}, \bibinfo {author} {\bibfnamefont {C.-Y.}\ \bibnamefont
  {Chen}}, \ and\ \bibinfo {author} {\bibfnamefont {P.}~\bibnamefont {Chen}},\
  }\href {\doibase 10.1140/epjc/s10052-014-2802-x} {\bibfield  {journal}
  {\bibinfo  {journal} {Eur. Phys. J.}\ }\textbf {\bibinfo {volume} {C74}},\
  \bibinfo {pages} {2802} (\bibinfo {year} {2014})},\ \Eprint
  {http://arxiv.org/abs/1302.5013} {arXiv:1302.5013 [gr-qc]} \BibitemShut
  {NoStop}%
\bibitem [{\citenamefont {Cho}\ and\ \citenamefont {Kim}(2014)}]{Cho:2014ija}%
  \BibitemOpen
  \bibfield  {author} {\bibinfo {author} {\bibfnamefont {I.}~\bibnamefont
  {Cho}}\ and\ \bibinfo {author} {\bibfnamefont {H.-C.}\ \bibnamefont {Kim}},\
  }\href {\doibase 10.1103/PhysRevD.90.024063} {\bibfield  {journal} {\bibinfo
  {journal} {Phys. Rev.}\ }\textbf {\bibinfo {volume} {D90}},\ \bibinfo {pages}
  {024063} (\bibinfo {year} {2014})},\ \Eprint {http://arxiv.org/abs/1404.6081}
  {arXiv:1404.6081 [gr-qc]} \BibitemShut {NoStop}%
\bibitem [{\citenamefont {Cho}\ and\ \citenamefont
  {Singh}(2014)}]{Cho:2014jta}%
  \BibitemOpen
  \bibfield  {author} {\bibinfo {author} {\bibfnamefont {I.}~\bibnamefont
  {Cho}}\ and\ \bibinfo {author} {\bibfnamefont {N.~K.}\ \bibnamefont
  {Singh}},\ }\href {\doibase 10.1140/epjc/s10052-014-3155-1} {\bibfield
  {journal} {\bibinfo  {journal} {Eur. Phys. J.}\ }\textbf {\bibinfo {volume}
  {C74}},\ \bibinfo {pages} {3155} (\bibinfo {year} {2014})},\ \Eprint
  {http://arxiv.org/abs/1408.2652} {arXiv:1408.2652 [gr-qc]} \BibitemShut
  {NoStop}%
\bibitem [{\citenamefont {Cho}\ and\ \citenamefont
  {Singh}(2015)}]{Cho:2014xaa}%
  \BibitemOpen
  \bibfield  {author} {\bibinfo {author} {\bibfnamefont {I.}~\bibnamefont
  {Cho}}\ and\ \bibinfo {author} {\bibfnamefont {N.~K.}\ \bibnamefont
  {Singh}},\ }\href {\doibase 10.1140/epjc/s10052-015-3458-x} {\bibfield
  {journal} {\bibinfo  {journal} {Eur. Phys. J.}\ }\textbf {\bibinfo {volume}
  {C75}},\ \bibinfo {pages} {240} (\bibinfo {year} {2015})},\ \Eprint
  {http://arxiv.org/abs/1412.6344} {arXiv:1412.6344 [gr-qc]} \BibitemShut
  {NoStop}%
\bibitem [{\citenamefont {Beltran~Jimenez}\ \emph {et~al.}(2015)\citenamefont
  {Beltran~Jimenez}, \citenamefont {Heisenberg}, \citenamefont {Olmo},\ and\
  \citenamefont {Ringeval}}]{Jimenez:2015jqa}%
  \BibitemOpen
  \bibfield  {author} {\bibinfo {author} {\bibfnamefont {J.}~\bibnamefont
  {Beltran~Jimenez}}, \bibinfo {author} {\bibfnamefont {L.}~\bibnamefont
  {Heisenberg}}, \bibinfo {author} {\bibfnamefont {G.~J.}\ \bibnamefont
  {Olmo}}, \ and\ \bibinfo {author} {\bibfnamefont {C.}~\bibnamefont
  {Ringeval}},\ }\href {\doibase 10.1088/1475-7516/2015/11/046} {\bibfield
  {journal} {\bibinfo  {journal} {JCAP}\ }\textbf {\bibinfo {volume} {1511}},\
  \bibinfo {pages} {046} (\bibinfo {year} {2015})},\ \Eprint
  {http://arxiv.org/abs/1509.01188} {arXiv:1509.01188 [gr-qc]} \BibitemShut
  {NoStop}%
\bibitem [{\citenamefont {de~Rham}\ and\ \citenamefont
  {Gabadadze}(2010)}]{deRham:2010ik}%
  \BibitemOpen
  \bibfield  {author} {\bibinfo {author} {\bibfnamefont {C.}~\bibnamefont
  {de~Rham}}\ and\ \bibinfo {author} {\bibfnamefont {G.}~\bibnamefont
  {Gabadadze}},\ }\href {\doibase 10.1103/PhysRevD.82.044020} {\bibfield
  {journal} {\bibinfo  {journal} {Phys.Rev.}\ }\textbf {\bibinfo {volume}
  {D82}},\ \bibinfo {pages} {044020} (\bibinfo {year} {2010})},\ \Eprint
  {http://arxiv.org/abs/1007.0443} {arXiv:1007.0443 [hep-th]} \BibitemShut
  {NoStop}%
\bibitem [{\citenamefont {de~Rham}\ \emph
  {et~al.}(2011{\natexlab{b}})\citenamefont {de~Rham}, \citenamefont
  {Gabadadze},\ and\ \citenamefont {Tolley}}]{deRham:2010kj}%
  \BibitemOpen
  \bibfield  {author} {\bibinfo {author} {\bibfnamefont {C.}~\bibnamefont
  {de~Rham}}, \bibinfo {author} {\bibfnamefont {G.}~\bibnamefont {Gabadadze}},
  \ and\ \bibinfo {author} {\bibfnamefont {A.~J.}\ \bibnamefont {Tolley}},\
  }\href {\doibase 10.1103/PhysRevLett.106.231101} {\bibfield  {journal}
  {\bibinfo  {journal} {Phys.Rev.Lett.}\ }\textbf {\bibinfo {volume} {106}},\
  \bibinfo {pages} {231101} (\bibinfo {year} {2011}{\natexlab{b}})},\ \Eprint
  {http://arxiv.org/abs/1011.1232} {arXiv:1011.1232 [hep-th]} \BibitemShut
  {NoStop}%
\bibitem [{\citenamefont {Hassan}\ and\ \citenamefont
  {Rosen}(2012)}]{Hassan:2011zd}%
  \BibitemOpen
  \bibfield  {author} {\bibinfo {author} {\bibfnamefont {S.}~\bibnamefont
  {Hassan}}\ and\ \bibinfo {author} {\bibfnamefont {R.~A.}\ \bibnamefont
  {Rosen}},\ }\href {\doibase 10.1007/JHEP02(2012)126} {\bibfield  {journal}
  {\bibinfo  {journal} {JHEP}\ }\textbf {\bibinfo {volume} {1202}},\ \bibinfo
  {pages} {126} (\bibinfo {year} {2012})},\ \Eprint
  {http://arxiv.org/abs/1109.3515} {arXiv:1109.3515 [hep-th]} \BibitemShut
  {NoStop}%
\bibitem [{\citenamefont {Vollick}(2004)}]{Vollick:2003qp}%
  \BibitemOpen
  \bibfield  {author} {\bibinfo {author} {\bibfnamefont {D.~N.}\ \bibnamefont
  {Vollick}},\ }\href {\doibase 10.1103/PhysRevD.69.064030} {\bibfield
  {journal} {\bibinfo  {journal} {Phys. Rev.}\ }\textbf {\bibinfo {volume}
  {D69}},\ \bibinfo {pages} {064030} (\bibinfo {year} {2004})},\ \Eprint
  {http://arxiv.org/abs/gr-qc/0309101} {arXiv:gr-qc/0309101 [gr-qc]}
  \BibitemShut {NoStop}%
\bibitem [{\citenamefont {de~Rham}\ \emph
  {et~al.}(2014{\natexlab{b}})\citenamefont {de~Rham}, \citenamefont {Matas},\
  and\ \citenamefont {Tolley}}]{deRham:2013tfa}%
  \BibitemOpen
  \bibfield  {author} {\bibinfo {author} {\bibfnamefont {C.}~\bibnamefont
  {de~Rham}}, \bibinfo {author} {\bibfnamefont {A.}~\bibnamefont {Matas}}, \
  and\ \bibinfo {author} {\bibfnamefont {A.~J.}\ \bibnamefont {Tolley}},\
  }\href {\doibase 10.1088/0264-9381/31/16/165004} {\bibfield  {journal}
  {\bibinfo  {journal} {Class. Quant. Grav.}\ }\textbf {\bibinfo {volume}
  {31}},\ \bibinfo {pages} {165004} (\bibinfo {year} {2014}{\natexlab{b}})},\
  \Eprint {http://arxiv.org/abs/1311.6485} {arXiv:1311.6485 [hep-th]}
  \BibitemShut {NoStop}%
\bibitem [{\citenamefont {Noller}\ and\ \citenamefont
  {Melville}(2014)}]{Noller:2014sta}%
  \BibitemOpen
  \bibfield  {author} {\bibinfo {author} {\bibfnamefont {J.}~\bibnamefont
  {Noller}}\ and\ \bibinfo {author} {\bibfnamefont {S.}~\bibnamefont
  {Melville}},\ }\href@noop {} {\  (\bibinfo {year} {2014})},\ \Eprint
  {http://arxiv.org/abs/1408.5131} {arXiv:1408.5131 [hep-th]} \BibitemShut
  {NoStop}%
\bibitem [{\citenamefont {{Mukhanov}}\ \emph {et~al.}(1992)\citenamefont
  {{Mukhanov}}, \citenamefont {{Feldman}},\ and\ \citenamefont
  {{Brandenberger}}}]{1992PhR...215..203M}%
  \BibitemOpen
  \bibfield  {author} {\bibinfo {author} {\bibfnamefont {V.~F.}\ \bibnamefont
  {{Mukhanov}}}, \bibinfo {author} {\bibfnamefont {H.~A.}\ \bibnamefont
  {{Feldman}}}, \ and\ \bibinfo {author} {\bibfnamefont {R.~H.}\ \bibnamefont
  {{Brandenberger}}},\ }\href {\doibase 10.1016/0370-1573(92)90044-Z}
  {\bibfield  {journal} {\bibinfo  {journal} {Physics Reports}\ }\textbf
  {\bibinfo {volume} {215}},\ \bibinfo {pages} {203} (\bibinfo {year}
  {1992})}\BibitemShut {NoStop}%
\bibitem [{\citenamefont {Boulware}\ and\ \citenamefont
  {Deser}(1972)}]{Boulware:1973my}%
  \BibitemOpen
  \bibfield  {author} {\bibinfo {author} {\bibfnamefont {D.}~\bibnamefont
  {Boulware}}\ and\ \bibinfo {author} {\bibfnamefont {S.}~\bibnamefont
  {Deser}},\ }\href {\doibase 10.1103/PhysRevD.6.3368} {\bibfield  {journal}
  {\bibinfo  {journal} {Phys.Rev.}\ }\textbf {\bibinfo {volume} {D6}},\
  \bibinfo {pages} {3368} (\bibinfo {year} {1972})}\BibitemShut {NoStop}%
\bibitem [{\citenamefont {Arkani-Hamed}\ \emph {et~al.}(2003)\citenamefont
  {Arkani-Hamed}, \citenamefont {Georgi},\ and\ \citenamefont
  {Schwartz}}]{ArkaniHamed:2002sp}%
  \BibitemOpen
  \bibfield  {author} {\bibinfo {author} {\bibfnamefont {N.}~\bibnamefont
  {Arkani-Hamed}}, \bibinfo {author} {\bibfnamefont {H.}~\bibnamefont
  {Georgi}}, \ and\ \bibinfo {author} {\bibfnamefont {M.~D.}\ \bibnamefont
  {Schwartz}},\ }\href {\doibase 10.1016/S0003-4916(03)00068-X} {\bibfield
  {journal} {\bibinfo  {journal} {Annals Phys.}\ }\textbf {\bibinfo {volume}
  {305}},\ \bibinfo {pages} {96} (\bibinfo {year} {2003})},\ \Eprint
  {http://arxiv.org/abs/hep-th/0210184} {arXiv:hep-th/0210184 [hep-th]}
  \BibitemShut {NoStop}%
\bibitem [{\citenamefont {Creminelli}\ \emph {et~al.}(2005)\citenamefont
  {Creminelli}, \citenamefont {Nicolis}, \citenamefont {Papucci},\ and\
  \citenamefont {Trincherini}}]{Creminelli:2005qk}%
  \BibitemOpen
  \bibfield  {author} {\bibinfo {author} {\bibfnamefont {P.}~\bibnamefont
  {Creminelli}}, \bibinfo {author} {\bibfnamefont {A.}~\bibnamefont {Nicolis}},
  \bibinfo {author} {\bibfnamefont {M.}~\bibnamefont {Papucci}}, \ and\
  \bibinfo {author} {\bibfnamefont {E.}~\bibnamefont {Trincherini}},\ }\href
  {\doibase 10.1088/1126-6708/2005/09/003} {\bibfield  {journal} {\bibinfo
  {journal} {JHEP}\ }\textbf {\bibinfo {volume} {09}},\ \bibinfo {pages} {003}
  (\bibinfo {year} {2005})},\ \Eprint {http://arxiv.org/abs/hep-th/0505147}
  {arXiv:hep-th/0505147 [hep-th]} \BibitemShut {NoStop}%
\bibitem [{\citenamefont {Higuchi}(1987)}]{Higuchi:1986py}%
  \BibitemOpen
  \bibfield  {author} {\bibinfo {author} {\bibfnamefont {A.}~\bibnamefont
  {Higuchi}},\ }\href {\doibase 10.1016/0550-3213(87)90691-2} {\bibfield
  {journal} {\bibinfo  {journal} {Nucl. Phys.}\ }\textbf {\bibinfo {volume}
  {B282}},\ \bibinfo {pages} {397} (\bibinfo {year} {1987})}\BibitemShut
  {NoStop}%
\bibitem [{\citenamefont {Noller}\ \emph {et~al.}(2014)\citenamefont {Noller},
  \citenamefont {Scargill},\ and\ \citenamefont {Ferreira}}]{Noller:2013yja}%
  \BibitemOpen
  \bibfield  {author} {\bibinfo {author} {\bibfnamefont {J.}~\bibnamefont
  {Noller}}, \bibinfo {author} {\bibfnamefont {J.~H.~C.}\ \bibnamefont
  {Scargill}}, \ and\ \bibinfo {author} {\bibfnamefont {P.~G.}\ \bibnamefont
  {Ferreira}},\ }\href {\doibase 10.1088/1475-7516/2014/02/007} {\bibfield
  {journal} {\bibinfo  {journal} {JCAP}\ }\textbf {\bibinfo {volume} {1402}},\
  \bibinfo {pages} {007} (\bibinfo {year} {2014})},\ \Eprint
  {http://arxiv.org/abs/1311.7009} {arXiv:1311.7009 [hep-th]} \BibitemShut
  {NoStop}%
\bibitem [{\citenamefont {Baker}\ \emph {et~al.}(2011)\citenamefont {Baker},
  \citenamefont {Ferreira}, \citenamefont {Skordis},\ and\ \citenamefont
  {Zuntz}}]{Baker:2011jy}%
  \BibitemOpen
  \bibfield  {author} {\bibinfo {author} {\bibfnamefont {T.}~\bibnamefont
  {Baker}}, \bibinfo {author} {\bibfnamefont {P.~G.}\ \bibnamefont {Ferreira}},
  \bibinfo {author} {\bibfnamefont {C.}~\bibnamefont {Skordis}}, \ and\
  \bibinfo {author} {\bibfnamefont {J.}~\bibnamefont {Zuntz}},\ }\href
  {\doibase 10.1103/PhysRevD.84.124018} {\bibfield  {journal} {\bibinfo
  {journal} {Phys. Rev.}\ }\textbf {\bibinfo {volume} {D84}},\ \bibinfo {pages}
  {124018} (\bibinfo {year} {2011})},\ \Eprint {http://arxiv.org/abs/1107.0491}
  {arXiv:1107.0491 [astro-ph.CO]} \BibitemShut {NoStop}%
\bibitem [{\citenamefont {Hu}\ \emph {et~al.}(2014)\citenamefont {Hu},
  \citenamefont {Raveri}, \citenamefont {Frusciante},\ and\ \citenamefont
  {Silvestri}}]{Hu:2013twa}%
  \BibitemOpen
  \bibfield  {author} {\bibinfo {author} {\bibfnamefont {B.}~\bibnamefont
  {Hu}}, \bibinfo {author} {\bibfnamefont {M.}~\bibnamefont {Raveri}}, \bibinfo
  {author} {\bibfnamefont {N.}~\bibnamefont {Frusciante}}, \ and\ \bibinfo
  {author} {\bibfnamefont {A.}~\bibnamefont {Silvestri}},\ }\href {\doibase
  10.1103/PhysRevD.89.103530} {\bibfield  {journal} {\bibinfo  {journal} {Phys.
  Rev.}\ }\textbf {\bibinfo {volume} {D89}},\ \bibinfo {pages} {103530}
  (\bibinfo {year} {2014})},\ \Eprint {http://arxiv.org/abs/1312.5742}
  {arXiv:1312.5742 [astro-ph.CO]} \BibitemShut {NoStop}%
\bibitem [{\citenamefont {Zumalacarregui}\ \emph {et~al.}(2016)\citenamefont
  {Zumalacarregui}, \citenamefont {Bellini}, \citenamefont {Sawicki},\ and\
  \citenamefont {Lesgourgues}}]{Zumalacarregui:2016pph}%
  \BibitemOpen
  \bibfield  {author} {\bibinfo {author} {\bibfnamefont {M.}~\bibnamefont
  {Zumalacarregui}}, \bibinfo {author} {\bibfnamefont {E.}~\bibnamefont
  {Bellini}}, \bibinfo {author} {\bibfnamefont {I.}~\bibnamefont {Sawicki}}, \
  and\ \bibinfo {author} {\bibfnamefont {J.}~\bibnamefont {Lesgourgues}},\
  }\href@noop {} {\  (\bibinfo {year} {2016})},\ \Eprint
  {http://arxiv.org/abs/1605.06102} {arXiv:1605.06102 [astro-ph.CO]}
  \BibitemShut {NoStop}%
\bibitem [{\citenamefont {{Brax}}\ \emph {et~al.}(2012)\citenamefont {{Brax}},
  \citenamefont {{Davis}}, \citenamefont {{Li}}, \citenamefont {{Winther}},\
  and\ \citenamefont {{Zhao}}}]{2012JCAP...10..002B}%
  \BibitemOpen
  \bibfield  {author} {\bibinfo {author} {\bibfnamefont {P.}~\bibnamefont
  {{Brax}}}, \bibinfo {author} {\bibfnamefont {A.-C.}\ \bibnamefont {{Davis}}},
  \bibinfo {author} {\bibfnamefont {B.}~\bibnamefont {{Li}}}, \bibinfo {author}
  {\bibfnamefont {H.~A.}\ \bibnamefont {{Winther}}}, \ and\ \bibinfo {author}
  {\bibfnamefont {G.-B.}\ \bibnamefont {{Zhao}}},\ }\href {\doibase
  10.1088/1475-7516/2012/10/002} {\bibfield  {journal} {\bibinfo  {journal}
  {JCAP}\ }\textbf {\bibinfo {volume} {10}},\ \bibinfo {eid} {002} (\bibinfo
  {year} {2012})},\ \Eprint {http://arxiv.org/abs/1206.3568} {arXiv:1206.3568}
  \BibitemShut {NoStop}%
\bibitem [{\citenamefont {{Brax}}\ \emph {et~al.}(2013)\citenamefont {{Brax}},
  \citenamefont {{Davis}}, \citenamefont {{Li}}, \citenamefont {{Winther}},\
  and\ \citenamefont {{Zhao}}}]{2013JCAP...04..029B}%
  \BibitemOpen
  \bibfield  {author} {\bibinfo {author} {\bibfnamefont {P.}~\bibnamefont
  {{Brax}}}, \bibinfo {author} {\bibfnamefont {A.-C.}\ \bibnamefont {{Davis}}},
  \bibinfo {author} {\bibfnamefont {B.}~\bibnamefont {{Li}}}, \bibinfo {author}
  {\bibfnamefont {H.~A.}\ \bibnamefont {{Winther}}}, \ and\ \bibinfo {author}
  {\bibfnamefont {G.-B.}\ \bibnamefont {{Zhao}}},\ }\href {\doibase
  10.1088/1475-7516/2013/04/029} {\bibfield  {journal} {\bibinfo  {journal}
  {JCAP}\ }\textbf {\bibinfo {volume} {4}},\ \bibinfo {eid} {029} (\bibinfo
  {year} {2013})},\ \Eprint {http://arxiv.org/abs/1303.0007} {arXiv:1303.0007
  [astro-ph.CO]} \BibitemShut {NoStop}%
\bibitem [{\citenamefont {Schmidt}(2009)}]{Schmidt:2009sg}%
  \BibitemOpen
  \bibfield  {author} {\bibinfo {author} {\bibfnamefont {F.}~\bibnamefont
  {Schmidt}},\ }\href {\doibase 10.1103/PhysRevD.80.043001} {\bibfield
  {journal} {\bibinfo  {journal} {Phys. Rev.}\ }\textbf {\bibinfo {volume}
  {D80}},\ \bibinfo {pages} {043001} (\bibinfo {year} {2009})},\ \Eprint
  {http://arxiv.org/abs/0905.0858} {arXiv:0905.0858 [astro-ph.CO]} \BibitemShut
  {NoStop}%
\bibitem [{\citenamefont {Barreira}\ \emph {et~al.}(2013)\citenamefont
  {Barreira}, \citenamefont {Li}, \citenamefont {Hellwing}, \citenamefont
  {Baugh},\ and\ \citenamefont {Pascoli}}]{Barreira:2013eea}%
  \BibitemOpen
  \bibfield  {author} {\bibinfo {author} {\bibfnamefont {A.}~\bibnamefont
  {Barreira}}, \bibinfo {author} {\bibfnamefont {B.}~\bibnamefont {Li}},
  \bibinfo {author} {\bibfnamefont {W.~A.}\ \bibnamefont {Hellwing}}, \bibinfo
  {author} {\bibfnamefont {C.~M.}\ \bibnamefont {Baugh}}, \ and\ \bibinfo
  {author} {\bibfnamefont {S.}~\bibnamefont {Pascoli}},\ }\href {\doibase
  10.1088/1475-7516/2013/10/027} {\bibfield  {journal} {\bibinfo  {journal}
  {JCAP}\ }\textbf {\bibinfo {volume} {1310}},\ \bibinfo {pages} {027}
  (\bibinfo {year} {2013})},\ \Eprint {http://arxiv.org/abs/1306.3219}
  {arXiv:1306.3219 [astro-ph.CO]} \BibitemShut {NoStop}%
\bibitem [{\citenamefont {Li}\ \emph {et~al.}(2013)\citenamefont {Li},
  \citenamefont {Barreira}, \citenamefont {Baugh}, \citenamefont {Hellwing},
  \citenamefont {Koyama}, \citenamefont {Pascoli},\ and\ \citenamefont
  {Zhao}}]{Li:2013tda}%
  \BibitemOpen
  \bibfield  {author} {\bibinfo {author} {\bibfnamefont {B.}~\bibnamefont
  {Li}}, \bibinfo {author} {\bibfnamefont {A.}~\bibnamefont {Barreira}},
  \bibinfo {author} {\bibfnamefont {C.~M.}\ \bibnamefont {Baugh}}, \bibinfo
  {author} {\bibfnamefont {W.~A.}\ \bibnamefont {Hellwing}}, \bibinfo {author}
  {\bibfnamefont {K.}~\bibnamefont {Koyama}}, \bibinfo {author} {\bibfnamefont
  {S.}~\bibnamefont {Pascoli}}, \ and\ \bibinfo {author} {\bibfnamefont
  {G.-B.}\ \bibnamefont {Zhao}},\ }\href {\doibase
  10.1088/1475-7516/2013/11/012} {\bibfield  {journal} {\bibinfo  {journal}
  {JCAP}\ }\textbf {\bibinfo {volume} {1311}},\ \bibinfo {pages} {012}
  (\bibinfo {year} {2013})},\ \Eprint {http://arxiv.org/abs/1308.3491}
  {arXiv:1308.3491 [astro-ph.CO]} \BibitemShut {NoStop}%
\bibitem [{\citenamefont {Falck}\ \emph {et~al.}(2014)\citenamefont {Falck},
  \citenamefont {Koyama}, \citenamefont {Zhao},\ and\ \citenamefont
  {Li}}]{Falck:2014jwa}%
  \BibitemOpen
  \bibfield  {author} {\bibinfo {author} {\bibfnamefont {B.}~\bibnamefont
  {Falck}}, \bibinfo {author} {\bibfnamefont {K.}~\bibnamefont {Koyama}},
  \bibinfo {author} {\bibfnamefont {G.-b.}\ \bibnamefont {Zhao}}, \ and\
  \bibinfo {author} {\bibfnamefont {B.}~\bibnamefont {Li}},\ }\href {\doibase
  10.1088/1475-7516/2014/07/058} {\bibfield  {journal} {\bibinfo  {journal}
  {JCAP}\ }\textbf {\bibinfo {volume} {1407}},\ \bibinfo {pages} {058}
  (\bibinfo {year} {2014})},\ \Eprint {http://arxiv.org/abs/1404.2206}
  {arXiv:1404.2206 [astro-ph.CO]} \BibitemShut {NoStop}%
\bibitem [{\citenamefont {Winther}\ and\ \citenamefont
  {Ferreira}(2015)}]{Winther:2014cia}%
  \BibitemOpen
  \bibfield  {author} {\bibinfo {author} {\bibfnamefont {H.~A.}\ \bibnamefont
  {Winther}}\ and\ \bibinfo {author} {\bibfnamefont {P.~G.}\ \bibnamefont
  {Ferreira}},\ }\href {\doibase 10.1103/PhysRevD.91.123507} {\bibfield
  {journal} {\bibinfo  {journal} {Phys. Rev.}\ }\textbf {\bibinfo {volume}
  {D91}},\ \bibinfo {pages} {123507} (\bibinfo {year} {2015})},\ \Eprint
  {http://arxiv.org/abs/1403.6492} {arXiv:1403.6492 [astro-ph.CO]} \BibitemShut
  {NoStop}%
\bibitem [{\citenamefont {Alonso}\ \emph {et~al.}(2016)\citenamefont {Alonso},
  \citenamefont {Bellini}, \citenamefont {Ferreira},\ and\ \citenamefont
  {Zumalacarregui}}]{Alonso:2016suf}%
  \BibitemOpen
  \bibfield  {author} {\bibinfo {author} {\bibfnamefont {D.}~\bibnamefont
  {Alonso}}, \bibinfo {author} {\bibfnamefont {E.}~\bibnamefont {Bellini}},
  \bibinfo {author} {\bibfnamefont {P.~G.}\ \bibnamefont {Ferreira}}, \ and\
  \bibinfo {author} {\bibfnamefont {M.}~\bibnamefont {Zumalacarregui}},\
  }\href@noop {} {\  (\bibinfo {year} {2016})},\ \Eprint
  {http://arxiv.org/abs/1610.09290} {arXiv:1610.09290 [astro-ph.CO]}
  \BibitemShut {NoStop}%
\bibitem [{\citenamefont {Folkerts}\ \emph {et~al.}(2011)\citenamefont
  {Folkerts}, \citenamefont {Pritzel},\ and\ \citenamefont
  {Wintergerst}}]{Folkerts:2011ev}%
  \BibitemOpen
  \bibfield  {author} {\bibinfo {author} {\bibfnamefont {S.}~\bibnamefont
  {Folkerts}}, \bibinfo {author} {\bibfnamefont {A.}~\bibnamefont {Pritzel}}, \
  and\ \bibinfo {author} {\bibfnamefont {N.}~\bibnamefont {Wintergerst}},\
  }\href@noop {} {\  (\bibinfo {year} {2011})},\ \Eprint
  {http://arxiv.org/abs/1107.3157} {arXiv:1107.3157 [hep-th]} \BibitemShut
  {NoStop}%
\bibitem [{\citenamefont {Hinterbichler}(2013)}]{Hinterbichler:2013eza}%
  \BibitemOpen
  \bibfield  {author} {\bibinfo {author} {\bibfnamefont {K.}~\bibnamefont
  {Hinterbichler}},\ }\href {\doibase 10.1007/JHEP10(2013)102} {\bibfield
  {journal} {\bibinfo  {journal} {JHEP}\ }\textbf {\bibinfo {volume} {10}},\
  \bibinfo {pages} {102} (\bibinfo {year} {2013})},\ \Eprint
  {http://arxiv.org/abs/1305.7227} {arXiv:1305.7227 [hep-th]} \BibitemShut
  {NoStop}%
\bibitem [{\citenamefont {Kimura}\ and\ \citenamefont
  {Yamauchi}(2013)}]{Kimura:2013ika}%
  \BibitemOpen
  \bibfield  {author} {\bibinfo {author} {\bibfnamefont {R.}~\bibnamefont
  {Kimura}}\ and\ \bibinfo {author} {\bibfnamefont {D.}~\bibnamefont
  {Yamauchi}},\ }\href {\doibase 10.1103/PhysRevD.88.084025} {\bibfield
  {journal} {\bibinfo  {journal} {Phys. Rev.}\ }\textbf {\bibinfo {volume}
  {D88}},\ \bibinfo {pages} {084025} (\bibinfo {year} {2013})},\ \Eprint
  {http://arxiv.org/abs/1308.0523} {arXiv:1308.0523 [gr-qc]} \BibitemShut
  {NoStop}%
\bibitem [{\citenamefont {Rubakov}(2004)}]{Rubakov:2004eb}%
  \BibitemOpen
  \bibfield  {author} {\bibinfo {author} {\bibfnamefont {V.~A.}\ \bibnamefont
  {Rubakov}},\ }\href@noop {} {\  (\bibinfo {year} {2004})},\ \Eprint
  {http://arxiv.org/abs/hep-th/0407104} {arXiv:hep-th/0407104 [hep-th]}
  \BibitemShut {NoStop}%
\bibitem [{\citenamefont {Dubovsky}(2004)}]{Dubovsky:2004sg}%
  \BibitemOpen
  \bibfield  {author} {\bibinfo {author} {\bibfnamefont {S.~L.}\ \bibnamefont
  {Dubovsky}},\ }\href {\doibase 10.1088/1126-6708/2004/10/076} {\bibfield
  {journal} {\bibinfo  {journal} {JHEP}\ }\textbf {\bibinfo {volume} {10}},\
  \bibinfo {pages} {076} (\bibinfo {year} {2004})},\ \Eprint
  {http://arxiv.org/abs/hep-th/0409124} {arXiv:hep-th/0409124 [hep-th]}
  \BibitemShut {NoStop}%
\bibitem [{\citenamefont {Rubakov}\ and\ \citenamefont
  {Tinyakov}(2008)}]{Rubakov:2008nh}%
  \BibitemOpen
  \bibfield  {author} {\bibinfo {author} {\bibfnamefont {V.~A.}\ \bibnamefont
  {Rubakov}}\ and\ \bibinfo {author} {\bibfnamefont {P.~G.}\ \bibnamefont
  {Tinyakov}},\ }\href {\doibase 10.1070/PU2008v051n08ABEH006600} {\bibfield
  {journal} {\bibinfo  {journal} {Phys. Usp.}\ }\textbf {\bibinfo {volume}
  {51}},\ \bibinfo {pages} {759} (\bibinfo {year} {2008})},\ \Eprint
  {http://arxiv.org/abs/0802.4379} {arXiv:0802.4379 [hep-th]} \BibitemShut
  {NoStop}%
\bibitem [{\citenamefont {Blas}\ \emph {et~al.}(2009)\citenamefont {Blas},
  \citenamefont {Comelli}, \citenamefont {Nesti},\ and\ \citenamefont
  {Pilo}}]{Blas:2009my}%
  \BibitemOpen
  \bibfield  {author} {\bibinfo {author} {\bibfnamefont {D.}~\bibnamefont
  {Blas}}, \bibinfo {author} {\bibfnamefont {D.}~\bibnamefont {Comelli}},
  \bibinfo {author} {\bibfnamefont {F.}~\bibnamefont {Nesti}}, \ and\ \bibinfo
  {author} {\bibfnamefont {L.}~\bibnamefont {Pilo}},\ }\href {\doibase
  10.1103/PhysRevD.80.044025} {\bibfield  {journal} {\bibinfo  {journal} {Phys.
  Rev.}\ }\textbf {\bibinfo {volume} {D80}},\ \bibinfo {pages} {044025}
  (\bibinfo {year} {2009})},\ \Eprint {http://arxiv.org/abs/0905.1699}
  {arXiv:0905.1699 [hep-th]} \BibitemShut {NoStop}%
\end{thebibliography}%

\end{document}